\documentstyle[apjfonts,emulateapj]{article}

\begin{document}
\submitted{Preprint: The Astrophysical Journal (1998), in press}

\title{HST and VLA Observations of Seyfert 2 Galaxies: The
Relationship between Radio Ejecta and the Narrow Line Region$^1$}

\footnotetext[1]{Based on
observations with the NASA/ESA Hubble Space Telescope, obtained at the
Space Telescope Science Institute, which is operated by AURA, Inc.,
under NASA contract NAS 5-26555 and on observations made with the VLA
operated by NRAO.}

\author{Heino Falcke}
\affil{
Astronomy Department, University of Maryland, College Park,
MD 20742-2421 
\\Max-Planck-Institut f\"ur
Radioastronomie, Auf dem H\"ugel 69, D-53121 Bonn, Germany
(hfalcke@mpifr-bonn.mpg.de)
}

\author{\vskip3mm
Andrew S. Wilson\altaffilmark{2}}
\affil{Astronomy Department, University of Maryland, College Park,
MD 20742-2421 (wilson@astro.umd.edu)}
\altaffiltext{2}{Adjunct Astronomer, Space Telescope Science Institute}

\author{\vskip3mmChris Simpson}
\affil{Jet Propulsion Laboratory, MS 169--327, 4800 Oak Grove Drive,
Pasadena, CA 91109 (bart@fornax.jpl.nasa.gov)}

\begin{abstract}
We present HST/WFPC2 images, in narrow-band filters containing the
[\ion{O}{3}]~$\lambda$5007 and H$\alpha$+[\ion{N}{2}] emission-lines
and their adjacent continua, of a sample of seven Seyfert~2 galaxies
selected on the basis of possessing either extended emission-line
regions in ground-based observations or a hidden broad-line region in
polarized light. Six of the galaxies have also been observed with the
VLA to obtain radio maps with better quality and angular resolution
than those in the literature. We find detailed correspondences between
features in the radio and emission-line images that clearly indicate
strong interactions between the radio jets and the interstellar
medium. Such interactions play a major role in determining the
morphology of the NLR, as the radio jets sweep up and compress ambient
gas, producing ordered structures with enhanced surface brightness in
line emission.  In at least three galaxies, namely Mrk~573,
ESO~428$-$G14, and Mrk~34, and perhaps also NGC~7212, off-nuclear
radio lobes coincide with regions of low gaseous excitation (as
measured by the [\ion{O}{3}]/(H$\alpha$+[\ion{N}{2}]) ratio). In
Mrk~573 and NGC~4388, there is a clear trend for low brightness
ionized gas to be of higher excitation. Those results may be
understood if radio lobes and regions of high emission-line surface
brightness are associated with high gas densities, reducing the
ionization parameter.  [\ion{O}{3}]/(H$\alpha$+[\ion{N}{2}])
excitation maps reveal bi-polar structures which can be interpreted as
either the `ionization cones' expected in the unified scheme or
widening, self-excited gaseous outflows. Only NGC 4388 and Mrk~573
show a clearly defined, straight-edged ionization cone.
\end{abstract}


\section{Introduction}
The strong tendency for the circumnuclear emission-line and radio
morphologies to be aligned in Seyfert galaxies (e.g.~Unger et
al.~1987; Pogge 1988; and Haniff, Wilson \& Ward 1988) added weight to
the idea that the ionizing radiation escapes preferentially from the
active nucleus along the radio axis. When some of these structures were
revealed to be well-defined cones by virtue of the improved resolution
offered by the Hubble Space Telescope (HST; e.g.\ NGC~1068, Evans et
al.\ 1991; NGC~5728, Wilson et al.\ 1993; NGC~5643, Simpson et al.\
1997), such anisotropic escape of ionizing photons was confirmed. This
effect is most popularly explained by the presence of an optically
thick `obscuring torus' (Antonucci 1993), which is able to collimate
the intrinsically isotropic ionizing radiation (see, e.g.,
Storchi-Bergmann, Mulchaey \& Wilson 1993). In addition, a number of
galaxies display the `ionization cone' morphology when an excitation
map is made, e.g.\ in [\ion{O}{3}]/(H$\alpha$+[\ion{N}{2}]).

The close connection between the radio ejecta of Seyfert nuclei and
their narrow line regions (NLRs) initially became apparent from their
similar spatial extents and from strong correlations between radio
luminosities and [\ion{O}{3}]$\lambda$5007 luminosity and line width
(de Bruyn \& Wilson 1978; Wilson \& Willis 1980; Whittle 1985, 1992).
Spectroscopic studies of the NLR (Baldwin, Wilson \& Whittle 1987;
Whittle et al.\ 1988), have revealed that the kinematics of the gas
are often clearly affected by the radio jets. Such interactions could
play a role in determining the structure of the NLR within the region
ionized by the nucleus. In a handful of cases, HST has shown a clear
spatial correspondence between the radio and emission-line
distributions (e.g.~NGC~5929, Bower et al.~1994; Mrk~78, Capetti et
al.~1994, 1996; Mrk~1066, Bower et al.\ 1995; Mrk~3, Capetti et al.\
1996; ESO~428--G14, Falcke et al.\ 1996b), indicating that the radio
ejecta strongly perturb the ionized gas, at least in these objects. It
has also been suggested that the hot gas associated with the shocks
generated by the interaction between the radio ejecta and the ambient
medium is a significant source of ionizing radiation (e.g.~Dopita
1995; Dopita \& Sutherland 1995; Bicknell et al. 1997; see also
reviews in Morse, Raymond,
\& Wilson 1996 and Wilson 1996).

It is therefore of great importance to study more Seyfert galaxies at
the high spatial resolution which only HST can provide, to determine
whether the morphology of the narrow line region is determined by the
nuclear ionizing radiation or by the interaction of radio jets with
the interstellar medium, or by a combination of both. In this paper we
present images taken with the Wide Field and Planetary Camera 2
(WFPC2) of seven Seyfert~2 galaxies, selected on the basis of
possessing either extended emission-line regions (as seen in
ground-based images) or broad lines in polarized light. For each
galaxy we have images in the light of the [\ion{O}{3}]~$\lambda$5007
line and the H$\alpha$+[\ion{N}{2}]~$\lambda\lambda$6548,6583 blend, as
well as in adjacent continua. We also present new radio maps taken
with the Very Large Array (VLA), almost all in `A-configuration',
providing an angular resolution comparable with that of the HST
images. Taken together, these allow us to compare directly the
structures of the line-emitting gas and radio plasma on scales of tens
of parsecs.

We give the details of our observations in Section~2. The results are
presented on a galaxy by galaxy basis in Section~3, while in Section~4
we discuss these results and their implications for the role of the
radio jets in shaping the narrow line region. Some concluding
remarks are given in Section~5. Throughout this paper we adopt a Hubble constant
$H_0 = 75$\,km\,s$^{-1}$\,Mpc$^{-1}$.

\section{Observations and Data Reduction}
\subsection{HST Observations}

All galaxies in our sample were observed with the WFPC2 on board the
HST. Except for observations of H$\alpha$+[\ion{N}{2}] and the red
continuum of ESO~428$-$G14, we used the Linear-Ramp-Filters (LRFs) of
the WFPC2, which have position-dependent wavelengths of peak
transmission, to image objects in their redshifted emission lines and
their adjacent continua. This means that in order to observe a galaxy
at a specific wavelength with the LRFs, the object had to be placed at
a specific position within the field of view (FOV) of the WFPC2. In
most cases the image will therefore not be taken with the Planetary
Camera (PC) but rather with one of the three chips of the Wide Field
Camera (WF). The former has a higher resolution with 0\farcs0455/pixel
while the latter have 0\farcs1/pixel. The transmissions and passbands
of the LRFs are also position-dependent; the transmission ranges
between 40\% and 80\% and the passband is
$\Delta\lambda/\lambda\sim1.3\%$, with a roughly triangular
transmission profile. The passband and the run of the central
wavelength with position translate into an upper limit of 14\arcsec\
for the size of an emission-line region that can be imaged with the
LRFs.

For our observations the filters and central wavelengths were chosen
such that they were centered on the redshifted H$\alpha$
($\lambda$6563\AA{}) and [\ion{O}{3}] $\lambda5007$\AA{} (hereafter
referred to simply as [\ion{O}{3}]) emission-lines and their adjacent
continua, all observations being split into two integrations to allow
cosmic ray rejection. Due to the relatively large filter width, the
[\ion{N}{2}]$\lambda\lambda6548,6583$ contribution to the H$\alpha$
images is significant (typically between a third and one half),
depending on the local [\ion{N}{2}]/H$\alpha$-ratio and the position
on the chip.

The details of the observations are given in Table~1. All exposures
were bias- and dark-subtracted by the WFPC2 pipeline processing at the
Space Telescope Science Institute. Since flatfields were not available
for the LRFs, we used flatfields taken in nearby narrow-band filters
(F502N and F673N).  The exposures were then combined to remove cosmic
rays and corrected for geometric distortions, since line- and
continuum images were in general taken at different positions on the
chip. The images were then rotated to the cardinal orientation (north
up, and east to the left).

\begin{deluxetable}{lllllllll}
\tablecaption{HST Observation Log}
\label{hstlog}
\tablehead{
\colhead{Galaxy}&
\colhead{$z$}&
\colhead{Filter}&
\colhead{$\lambda_{\rm c}$}&
\colhead{Chip}&
\colhead{Exp.}&
\colhead{Date}&
\colhead{Filename}&
\colhead{ID}
}
\tablecolumns{9}
\startdata
Mrk~348&    0.01514 &FR533N   & 5080 & WF2 &     300.00 &  16Dec1995   & U2XI0601(2)T &[\ion{O}{3}] $\lambda5007$\\        
(NGC~262)&  &FR533P15 & 5330 & WF4 &     140.00 &              & U2XI0603(4)T &[\ion{O}{3}] cont.\\  
     &     &FR680N   & 6666 & WF2 &     300.00 &              & U2XI0605(6)T &H$\alpha$+[\ion{N}{2}]\\      
     &     &FR680N   & 6580 & WF3 &     140.00 &              & U2XI0607(8)T &H$\alpha$+[\ion{N}{2}] cont.\\
Mrk~573& 0.0173\hphantom{0}&FR533N   & 5092 & WF2 &     300.00 &  12Nov1995   & U2XI0701(2)T &[\ion{O}{3}] $\lambda5007$\\        
(UGC01214)&&FR533N   & 5345 & WF4 &     140.00 &              & U2XI0703(4)T &[\ion{O}{3}] cont.\\  
       &   &FR680N   & 6679 & WF2 &     300.00 &              & U2XI0705(6)T &H$\alpha$+[\ion{N}{2}]\\      
       &   &FR680N   & 6514 & WF3 &     140.00 &              & U2XI0707(8)T &H$\alpha$+[\ion{N}{2}] cont.\\
ESO~428$-$G14&0.00544&FR533N   & 5034 & WF2 &     140.00 &  14Jan1996   & U2XI0301(2)T &[\ion{O}{3}] $\lambda5007$\\ 
          &&FR533N   & 5284 & WF3 &     140.00 &              & U2XI0303(4)T &[\ion{O}{3}] cont. \\ 
          &&F658N    & 6590 & PC1 &     400.00 &  17Apr1995   & U2NP0601(2)T &H$\alpha$+[\ion{N}{2}]\\ 
          &&F814W    & 7940 & PC1 &     100.00 &              & U2NP0603(4)T &H$\alpha$+[\ion{N}{2}] cont.\\ 
Mrk~1210&  0.0135\hphantom{0}&FR533N   & 5068 & WF2 &     140.00 &  24Nov1995   & U2XI0501(2)T &[\ion{O}{3}] $\lambda5007$\\        
(UGC04203)&&FR533P15 & 5319 & WF4 &     140.00 &              & U2XI0503(4)T &[\ion{O}{3}] cont.\\  
       &   &FR680N   & 6654 & WF2 &     350.00 &              & U2XI0505(6)T &H$\alpha$+[\ion{N}{2}]\\      
       &   &FR680N   & 6551 & WF3 &     140.00 &              & U2XI0507(8)T &H$\alpha$+[\ion{N}{2}] cont.\\
Mrk~34&0.0505\hphantom{0}&FR533    & 5263 & WF3 &     230.00 &  08Nov1995   & U2XI0101(2)T &[\ion{O}{3}] $\lambda5007$\\
    &      &FR533N18 & 5499 & WF3 &     160.00 &              & U2XI0103(4)T &[\ion{O}{3}] cont.\\
    &      &FR680N18 & 6900 & WF2 &     140.00 &              & U2XI0105(6)T &H$\alpha$+[\ion{N}{2}]\\
    &      &FR680N   & 6727 & WF2 &     140.00 &              & U2XI0107(8)T &H$\alpha$+[\ion{N}{2}] cont.\\
NGC~4388&0.00842&FR533N   & 5053 & WF2 &     140.00 &  27Mar1996   & U2XI0201(2)T &[\ion{O}{3}] $\lambda5007$\\
      &    &FR533N   & 5304 & WF3 &     140.00 &              & U2XI0203(4)T &[\ion{O}{3}] cont.\\
      &    &FR680P15 & 6614 & PC1 &     350.00 &              & U2XI0205(6)T &H$\alpha$+[\ion{N}{2}]\\
      &    &FR680P   & 6448 & WF3 &     140.00 &              & U2XI0207(8)T &H$\alpha$+[\ion{N}{2}] cont.\\
NGC~7212&0.0266&FR533N   & 5136 & WF2 &     300.00 &  26Sep1995   & U2XI0401(2)T &[\ion{O}{3}] $\lambda5007$\\        
      &    &FR533N   & 5388 & WF4 &     140.00 &              & U2XI0403(4)T &[\ion{O}{3}] cont. \\ 
      &    &FR680N   & 6734 & WF2 &     346.50 &              & U2XI0405(6)T &H$\alpha$+[\ion{N}{2}]\\      
      &    &FR680N   & 6567 & WF3 &     140.00 &              & U2XI0407(8)T &H$\alpha$+[\ion{N}{2}] cont.\\
\tablecomments{Description of columns: 
(1) -- galaxy,
(2) -- redshift,
(3) -- filter name,
(4) -- central wavelength in \AA{}ngstroms,
(5) -- WFPC2 chip,
(6) -- exposure time of each exposure in seconds (total exposure time of each image is twice that value),
(7) -- observation date,
(8) -- root filename(s) of the two (CR-split) exposures per image,
(9) -- identification for this paper
}
\enddata
\end{deluxetable}

The continuum images were subtracted from the line images by shifting
them to a common reference frame according to the coordinate
information in the image headers. Usually each of the on- and off-band
pairs were taken during the same orbit with guide stars in fine lock,
hence the relative astrometry should be better than 20 mas. However,
since all our observations were interrupted by earth occultations we
could not rely in all cases solely on the spacecraft information to
align the [\ion{O}{3}] and H$\alpha$+[\ion{N}{2}] images of each
galaxy. We therefore aligned the green and red images according to
their continuum peaks and checked the results by cross-correlating
bright features in the emission-line images. With this method we
achieved sub-pixel (i.e. $\sim$0.5 pixel) precision in all cases for
the relative alignment of the continuum-subtracted emission-line
images.  This additional step is essential in order to avoid artifacts
in the excitation maps.

\subsection{VLA Observations}
In order to compare the HST images with radio maps of similar
resolution we have obtained new radio observations of all galaxies in
our sample with the VLA except for Mrk~348, which has a known VLBI
core (Neff \& de Bruyn 1983). All galaxies were observed at X-band
(3.5 cm) in `A-configuration', except for NGC~4388 which was observed
in `BnA-configuration'. In addition, U-band (2 cm) observations were
obtained for Mrk~573 and ESO~428$-$G14, and C-band (6.1 cm)
observations for Mrk~34. The best resolutions achieved were 0\farcs14,
0\farcs24, and 0\farcs35 at 2, 3.5, and 6.1 cm respectively.  For
scheduling reasons two of the sources (NGC~7212 and Mrk~1210) were
observed in snapshot mode only. The details of these observations are
given in Table~2.

\begin{deluxetable}{llllll}
\tablecaption{VLA Observation Log}
\label{vlalog}
\tablehead{
\colhead{Galaxy}&
\colhead{Config}&
\colhead{Band}&
\colhead{Int.~Time}&
\colhead{Calibrator}&
\colhead{Date}
}
\tablecolumns{6}
\startdata
Mrk~573     & A   & 2.0 cm & 3:20 & 0122$-$003 &  04 Nov 1996\\
          & A   & 3.5 cm & 1:04 & 0122$-$003 &  04 Nov 1996\\
ESO~428$-$G14& A   & 2.0 cm & 2:56 & 0646$-$306 &  04 Nov 1996\\
          & A   & 3.5 cm & 1:05 & 0646$-$306 &  04 Nov 1996\\
Mrk~1210   & A   & 3.5 cm & 0:15 & 0745+100 &  04 Nov 1996\\
Mrk~34      & A   & 3.5 cm & 2:42 & 1030+611 &  04 Nov 1996\\
          & A   & 6.1 cm & 1:13 & 1030+611 &  04 Nov 1996\\
NGC~4388   & BnA & 3.5 cm & 1:11 & 1239+075 &  01 Feb 1997\\
NGC~7212   & A   & 3.5 cm & 0:14 & 2144+092 &  04 Nov 1996\\
\tablecomments{
(1) -- galaxy,
(2) -- VLA array configuration,
(3) -- observed wavelength, 
(4) -- total integration time on source (h:mm),
(5) -- phase calibrator,
(6) -- observing date.
}
\enddata
\end{deluxetable}

Weather conditions were excellent for all observations. Integrations
on the galaxies were sandwiched between observations of a calibrator
with typical on-galaxy integrations of 8 and 12 mins at 2 cm and 3.5 cm
respectively.  Data reduction was done in Socorro with AIPS following
standard procedures with very little flagging of bad data points being
necessary. The phases of the calibrators were interpolated and applied
to the program objects and 3C286 was used as the primary flux
calibrator.  Maps of the galaxies were produced using the AIPS task
{\sc IMAGR} with various beam sizes, cleaning depths of several
thousand iterations, a CLEAN gain factor of 0.1, and uniform weighting
for high-resolution maps and natural weighting for
low-resolution maps.

\subsection{Radio and optical registration}\label{registration}
Since absolute positions derived from HST observations have an error
of the order 1\arcsec, the registration between the radio and the
optical frames is a critical issue.  Meaningful registration cannot,
therefore, rely on the internal HST astrometry and one has to shift
the HST images to align with the radio. Unfortunately, with the
exception of NGC~4388 and NGC~7212 (see below), ground-based
astrometric measurements are not yet available for stars in the WFPC2
frames, so we have used the following procedure for aligning the
images: first, we smoothed the red continuum HST image to a resolution
of $1\farcs5$ (FWHM), thus simulating a ground-based, seeing-limited
image. The red continuum was chosen because it generally has a higher
signal-to-noise ratio and we did not find significant offsets between
the green and the red peaks.  We then determined the peak in the
smoothed HST image, assigned to it the coordinates which were
previously determined for those galaxies by ground-based astrometry
(Columns~2 and 3 in Table~3), in most cases with a precision of tenths
of an arcsecond, and used this as the new coordinate system for the
HST images. The resulting shifts of the HST coordinates varied between
0\farcs5 and 2\arcsec. With the exception of two galaxies (NGC~4388
and NGC~7212) the exact width of the Gaussian used for smoothing
(``seeing'') did not make a significant difference to the position of
the peak in the smoothed HST image.

In Columns~4 \& 5 of Table~3, we give the positions of the
continuum peaks of the unsmoothed HST image in the corrected
(i.e.~ground-based) coordinate frame. Comparison of Columns~2 \& 3 with
Columns~4 \& 5 shows that in general the peaks in the smoothed and
unsmoothed images (both measured in the same frame) agree with each
other to better than two WF pixels (0\farcs2). For two objects
(ESO~428$-$G14 and Mrk~1210), no  subarcsecond accuracy
positions are available yet and we have listed the best available
positions from the literature.

\begin{deluxetable}{lllllll}
\tablecaption{Optical and Radio Positions}
\label{pos}
\tablehead{
\colhead{galaxy}&
\multicolumn{2}{c}{optical position}&  
\multicolumn{2}{c}{HST nucleus}&  
\multicolumn{2}{c}{VLA nucleus}\\
&
\colhead{h m s}&
\colhead{$^\circ$ \arcmin\ \arcsec}&
\colhead{h m s}&
\colhead{$^\circ$ \arcmin\ \arcsec}&
\colhead{h m s}&
\colhead{$^\circ$ \arcmin\ \arcsec}\\
& 
\multicolumn{2}{c}{J2000/B1950}&
\multicolumn{2}{c}{J2000/B1950}&
\multicolumn{2}{c}{J2000/B1950}\\
\colhead{(1)}&
\colhead{(2)}&
\colhead{(3)}&
\colhead{(4)}&
\colhead{(5)}&
\colhead{(6)}&
\colhead{(7)}
}
\tablecolumns{7}
\startdata
Mrk~348    & 00 48 47.128 &$+$31 57 25.45$^a$& 00 48 47.130 &$+$31 57 25.44 & 00 48 47.107 &$+$31 57 25.32$^b$\\
           & 00 46 04.882 &$+$31 41 04.58    & 00 46 04.884 &$+$31 41 04.57 & 00 46 04.861 &$+$31 41 04.45\\
Mrk~573     & 01 43 57.772 &$+$02 20 59.67$^a$& 01 43 57.771 &$+$02 20 59.66 & 01 43 57.748 &$+$02 20 59.58$^h$\\
           & 01 41 22.922 &$+$02 05 56.32    & 01 41 22.921 &$+$02 05 56.30 & 01 41 22.898 &$+$02 05 56.22\\
ESO~428$-$G14 & 07 16 31.2  &$-$29 19 28$^c$& 07 16 31.2 &$-$29 19 28 & 07 16 31.207 &$-$29 19 28.89$^h$\\
           & 07 14 32.9   &$-$29 14 04       & 07 14 32.9 &$-$29 14 04 & 07 14 32.930 &$-$29 14 04.51\\
Mrk~1210    & 08 04 05.95  &$+$05 06 50.4 $^d$& 08 04 05.95 &$+$05 06 50.4 & 08 04 05.841 &$+$05 06 49.72$^h$\\
           & 08 01 27.01  &$+$05 15 22.0     & 08 01 27.01 &$+$05 15 22.0 & 8 01 26.902  &$+$05 15 21.32\\
Mrk~34      & 10 34 08.583 &$+$60 01 51.74$^a$& 10 34 08.593 &$+$60 01 51.62 & 10 34 08.568 &$+$60 01 51.82$^h$\\
           & 10 30 51.458 &$+$60 17 21.84    & 10 30 51.468 &$+$60 17 21.72 & 10 30 51.443 &$+$60 17 21.92\\
NGC~4388    & 12 25 46.740 &$+$12 39 42.05$^e$ & 12 25 46.715 &$+$12 39 43.41$^e$ & 12 25 46.745 &$+$12 39 43.51$^f$\\
           & 12 23 14.635 &$+$12 56 18.58    & 12 23 14.610 &$+$12 56 19.94 & 12 23 14.640 &$+$12 56 20.04\\
NGC~7212    & 22 07 02.006 &$+$10 14 00.75$^g$& 22 07 01.998 &$+$10 14 00.75 & 22 07 01.934 &$+$10 14 01.01$^h$\\
           & 22 04 33.976 &$+$09 59 20.36    & 22 04 33.967 &$+$09 59 20.37 & 22 04 33.903 &$+$09 59 20.63

\tablecomments{
(1) -- name of galaxy;
(2 \& 3) -- RA and DEC of the optical nucleus as measured from the
ground (upper is in J2000, lower in B1950);
(4 \& 5) -- RA and DEC of the red HST continuum peak from the ground
based astrometric positions in cols.~2 \& 3 (see text);
(6 \& 7) -- RA and DEC of the VLA radio component identified as the nucleus.\\
 References for ground based positions and uncertainties: \\
($^a$) Clements (1981) -- $\sigma_{\rm RA/DEC}\simeq0\farcs15$;\\
($^b$) Ulvestad \& Wilson (1984) -- $\sigma_{\rm RA/DEC}\simeq0\farcs1$;\\
($^c$) V\'eron-Cetty \& V\'eron (1996) -- $\sigma_{\rm RA/DEC}\simeq2\farcs5$;\\
($^d$) Kojoian et al. (1981) -- $\sigma_{\rm RA/DEC}\simeq1\farcs5$;\\
($^e$) the optical coordinates for the nucleus of this galaxy were
derived from the HST images which were registered by reference 
to a nearby field star (12 25 49.869 +12 40 47.95 J2000) which
happened  to be visible in the HST FOV. The position of this star
relative to the HIPPARCOS frame was provided by R.W.~Argyle
(priv.~comm.) and is accurate to within 0\farcs1 in RA and DEC. The derived
HST positions have errors of $\sigma_{\rm RA/DEC}\simeq0\farcs2$;\\
($^f$) the position of the radio nucleus was taken from the VLA 2 cm
A-configuration observations of Carral et al.~(1990);\\
($^g$) optical coordinates for the nucleus of this galaxy were derived
similar to NGC~4388, the reference field star was at (J2000) 22 06
58.240 10 15 05.01, (R.W. Argyle, priv. comm) and the estimated 
errors of the HST positions are $\sigma_{\rm RA/DEC}\simeq0\farcs2$;
($^h$) this paper with an error better than 0\farcs3.
}
\enddata
\end{deluxetable}

Even though NGC~4388 has astrometric positions of sub-arcsecond
accuracy in the literature (Argyle \& Eldridge 1990, hereafter AE90),
we found this position to be useless for our purposes: because of the
asymmetric brightness distribution of the HST continuum image of this
galaxy, the position of the peak in the smoothed HST continuum image
depends critically on the assumed width of the Gaussian smoothing
function. We also found a large offset between the radio and the
optical positions. For NGC~7212 we also observe a large offset between
radio and the astrometric optical position (Clements 1983) and
hence needed to check this registration as well.

R.W.~Argyle (priv. comm.) kindly provided us with astrometric
positions of two field stars visible on the HST images near NGC~4388
and NGC~7212, which we then used to assign an absolute coordinate system to
the HST frames (see notes to Table~3). By comparing the positions of
the same stars in our FOV in different exposures at different
wavelengths (i.e. different positions in the LRF images) we find that
the accuracy of those positions should be better than $\sim0\farcs2$
accross the FOV of the WFC.

Moreover, we found that our own position for the radio nucleus of
NGC~4388 (which was observed with a very elongated beam with the VLA
BnA-configuration) differed by 0\farcs5 in RA and 0\farcs2 in DEC from
the positions given in Hummel \& Saikia (1991; observed at 6 cm with a
1\farcs2 beam) and in Carral et al.~(1990; observed at 2 cm with a
0\farcs2 beam). Since the positions from the latter two observations
agree very well with each other, while bridging our observations in
frequency and resolution, we concluded that our position was in error
and hence used the position of the radio nucleus given by Carral et
al.~(1990; see Table 3) for the radio/optical registration.

For those galaxies with ground-based optical positions of
sub-arcsecond accuracy, the agreement between the radio and the
optical frame should be better than $\sim$0\farcs4 (see Fig.~\ref{f1}
in AE90). With the exception of NGC~7212, the radio and optical RAs
and DECs do agree to within this accuracy (compare columns (4), (5),
(6), and (7) of Table 3). However, this is still not good enough,
given the $\simeq$0\farcs1 resolution of the HST images.  To obtain an
even better, though somewhat arbitrary, alignment, we applied a
further shift for each galaxy, as discussed on a galaxy by galaxy
basis below and in Section \ref{comments}. In general, we tried to
identify the nucleus in our radio images (usually the blob which was
closest to the nominal optical position) and then aligned the radio
nucleus with the optical continuum peak.  This leaves, of course, the
inherent uncertainty of whether the radio nucleus is properly
identified and whether there might be an offset between the radio and
optical nuclei due to, for example, obscuration. This issue can only
be solved by accurate, absolute optical astrometry of field stars
visible in the HST frames or by exact IR astrometry with large-format
arrays.

For Mrk~34 and Mrk~573 we find that the central radio peak and the
optical peak (derived from the astrometric measurements) agree to within
0\farcs3, and so we shifted the HST nucleus onto the radio
position. However, in NGC~7212 and NGC~4388 the difference between
the optical peak (in the astrometric frame) and the radio peak seems larger
than the expected errors and we did not perform a shift.

For the two objects which do not have optical positions of
sub-arcsecond accuracy, we registered optical emission-line and radio
images by their morphological similarity. Smoothing of radio and
optical emission-line structures to the same resolution yields a
relatively robust alignment to within 0\farcs3 for ESO~428$-$G14
(cf.~Falcke et al.~1996b). The registration of Mrk~1210 is even
simpler, since both optical and radio images are basically point
sources, which we assumed are coincident.

\section{Results}
Our HST observations of ESO 428$-$G14 and Mrk~573 have been, or will
be presented elsewhere (Falcke et al.~1996b; Wilson et al. 1998) and
we show here only their overlays on the new radio maps. Images of the
remaining galaxies in our sample are shown in
Figures~\ref{f1}-\ref{f5}. For each galaxy except Mrk~1210 we show the
continuum subtracted H$\alpha$+[\ion{N}{2}] and [\ion{O}{3}] images,
the excitation map (i.e. the [\ion{O}{3}] images divided by the
H$\alpha$+[\ion{N}{2}] image --- both were clipped below their
3$\sigma$ noise levels prior to the division), and a red continuum
image --- either our own narrow-band continuum image or, if available,
a broad-band snapshot image from the HST archive (from the snapshot survey of
galaxies by Malkan et al. 1997).  The emission-line images of Mrk~1210
are dominated by a single nuclear peak, so the excitation map is
omitted. We have compared our large aperture fluxes with those
obtained from ground based observations as compiled in Mulchaey et
al.~(1994) and find general agreement to within 20\%. Only for
NGC~4388 is the discrepancy larger, an effect most likely due to the
large size of its NLR. The fluxes of H$\alpha$+[\ion{N}{2}] and
[\ion{O}{3}] $\lambda5007$ for two different aperture sizes are given
in Table~4.

\begin{deluxetable}{lcccccc}
\tablecaption{HST Emission-Line Fluxes}
\label{fluxes}
\tablehead{
\colhead{Galaxy}&
\colhead{Aperture 1}&  
\colhead{H$\alpha$+[\ion{N}{2}] flux}&  
\colhead{[\ion{O}{3}] $\lambda5007$ flux}&
\colhead{Aperture 2}&  
\colhead{H$\alpha$+[\ion{N}{2}] flux}&  
\colhead{[\ion{O}{3}] $\lambda5007$ flux}\\
\colhead{}&
\colhead{[\arcsec]}&
\multicolumn{2}{c}{[$\times10^{-13}$erg\,sec$^{-1}$\, cm$^{-2}$]}&
\colhead{[\arcsec]}&
\multicolumn{2}{c}{[$\times10^{-13}$erg\,sec$^{-1}$\, cm$^{-2}$]}\\
\colhead{(1)}&
\colhead{(2)}&
\colhead{(3)}&
\colhead{(4)}&
\colhead{(5)}&
\colhead{(6)}&
\colhead{(7)}
}
\tablecolumns{7}
\startdata
Mrk~348     &  3 &  2.0 &  4.1 & 0.8 & 1.5 & 3.2 \\
Mrk~573     & 10 &  7.1 & 14.3 & 1.0 & 1.5 & 3.5 \\
ESO~428$-$G14  &  6 &  9.6 & 17.4 & 1.0 & 2.0 & 4.0 \\
Mrk~1210    &  2 &  3.8 & 10.4 & 0.6 & 2.9 & 8.9 \\
Mrk~34      &  4 &  3.6 &  6.1 & 1.0 & 0.7 & 1.3 \\
NGC~4388    & 12 & 10.1 & 11.7 & 2.0 & 1.9 & 3.0 \\
NGC~7212    &  6 &  4.5 &  9.3 & 0.8 & 1.7 & 3.9 
\tablecomments{The fluxes given here were derived by summing the flux
in the HST images inside a circular aperture centered on the
optical continuum nucleus. The apertures were chosen such that the first
aperture includes most of the NLR  visible in the HST images and
the second aperture is intended to represent the nuclear region. Description of the
columns:
(1) -- name of galaxy,
(2) -- diameter of large aperture in arcseconds,
(3 \& 4) -- emission-line fluxes in the H$\alpha$+[\ion{N}{2}] and [\ion{O}{3}] $\lambda5007$ filters for the large aperture,
(5) -- diameter of small aperture in arcseconds,
(6 \& 7) -- emission-line fluxes in the H$\alpha$+[\ion{N}{2}]  and [\ion{O}{3}] $\lambda5007$ filters for the small aperture.
}
\enddata
\end{deluxetable}

\begin{figure*}
\plotone{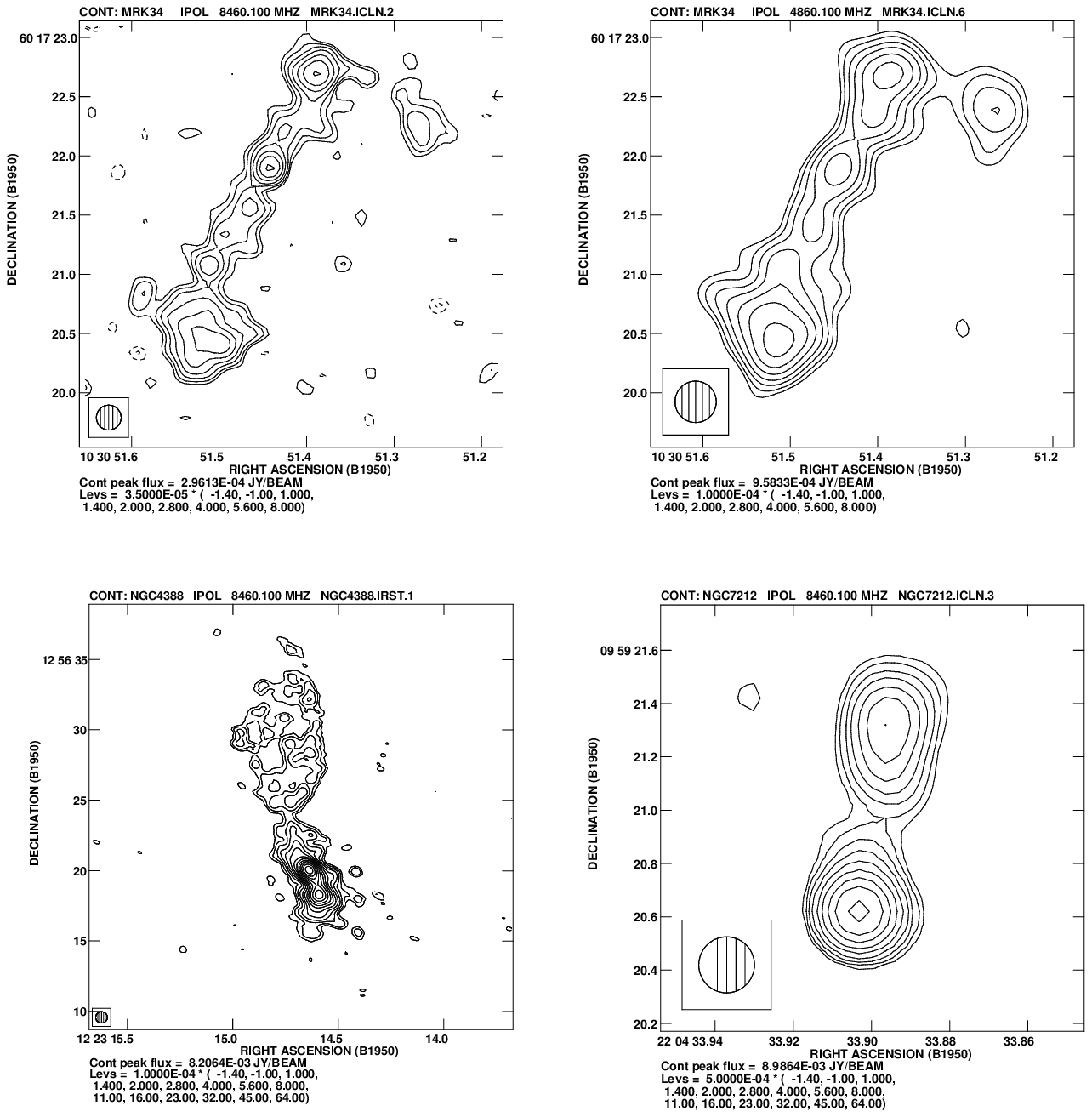}
\caption[]{\label{f8}VLA maps with B1950 coordinates. North is up, east is to the left. 
Upper left: Mrk~34 at 3.5 cm with 0\farcs21 beam.
The peak flux is 0.30 mJy (beam area)$^{-1}$ and contours
are plotted at -1.4, -1 (dotted), 1, 1.4, 2, 2.8, 4, 5.6 and 8.0
times 0.035 mJy (beam area)$^{-1}$.
Upper right: Mrk~34 at 6.1 cm with 0\farcs35 beam.
The peak flux is 0.96 mJy (beam area)$^{-1}$ and contours
are plotted at  -1.4, -1 (dotted), 1, 1.4, 2, 2.8, 4, 5.6 and 8.0
times 0.1  mJy (beam area)$^{-1}$.
Lower left: NGC~4388 at 3.5 cm with 0\farcs8 beam.
The peak flux is 8.2 mJy (beam area)$^{-1}$ and contours
are plotted at -1.4, -1 (dotted), 1, 1.4, 2, 2.8, 4, 5.6, 8.0,
11.0, 16.0, 23.0, 32.0, 45.0, and 64.0 
times 0.1 mJy (beam area)$^{-1}$.
The coordinates for our map of NGC~4388 differ slightly from those of
Hummel
\& Saikia (1991) and Carral et al.~(1990) (see Sect. 2.3).
Lower right: NGC~7212 at 3.5 cm with 0\farcs21 beam. 
The peak flux is 9.0 mJy (beam area)$^{-1}$ and contours
are plotted at -1.4, -1 (dotted), 1, 1.4, 2, 2.8, 4, 5.6, 8.0,
11.0, and 16 times 0.5  mJy (beam area)$^{-1}$.
}\end{figure*}

\begin{figure*}
\plotone{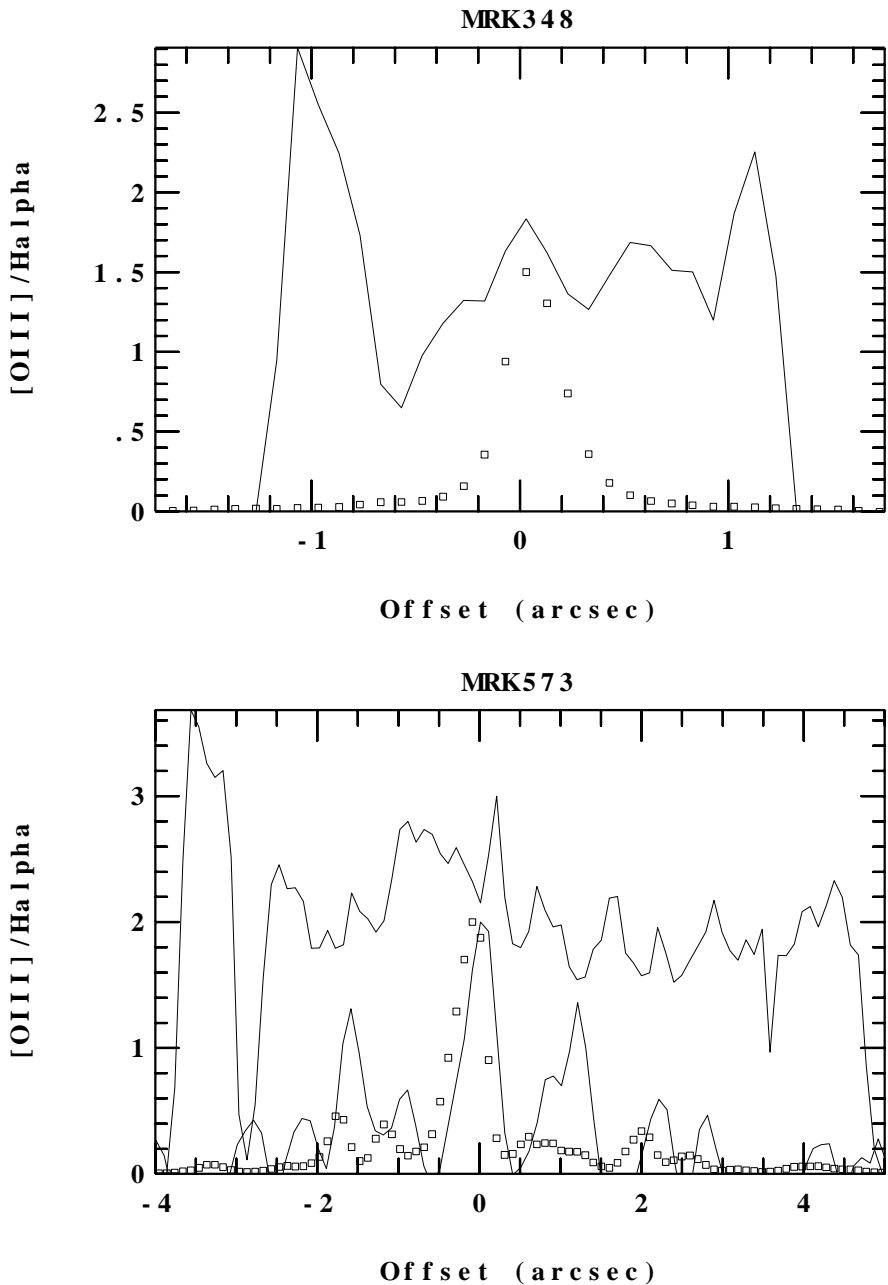}
\caption[]{\label{f9}Three pixel wide cross sections along the position angle of
the radio structure through the radio nucleus which is at zero offset
(north and NW directions have positive offsets). Top solid line --
[\ion{O}{3}]/(H$\alpha$+[\ion{N}{2}]) ratio, boxes --
H$\alpha$+[\ion{N}{2}] flux, bottom solid line -- radio flux at the
highest available frequency.
The ordinate scale applies to both the [\ion{O}{3}]/(H$\alpha$+[\ion{N}{2}])
ratios and the H$\alpha$+[\ion{N}{2}] brightnesses; the units for the
latter are given below.
Top panel: Mrk~348 at PA=0$^\circ$ (no radio flux
distribution); the ordinate scale for the H$\alpha$+[\ion{N}{2}]
brightness is in units of
$3.2\times10^{-15}$ erg sec$^{-1}$ cm$^{-2}$ 0\farcs1$^{-2}$.  Bottom
panel: Mrk~573 at PA = $-$54$^\circ$ (2 cm map); the ordinate scale for
the H$\alpha$+[\ion{N}{2}] brightness is in
units of $2.95\times10^{-15}$ erg sec$^{-1}$
cm$^{-2}$ 0\farcs1$^{-2}$.
}\end{figure*}

\begin{figure*}
\plotone{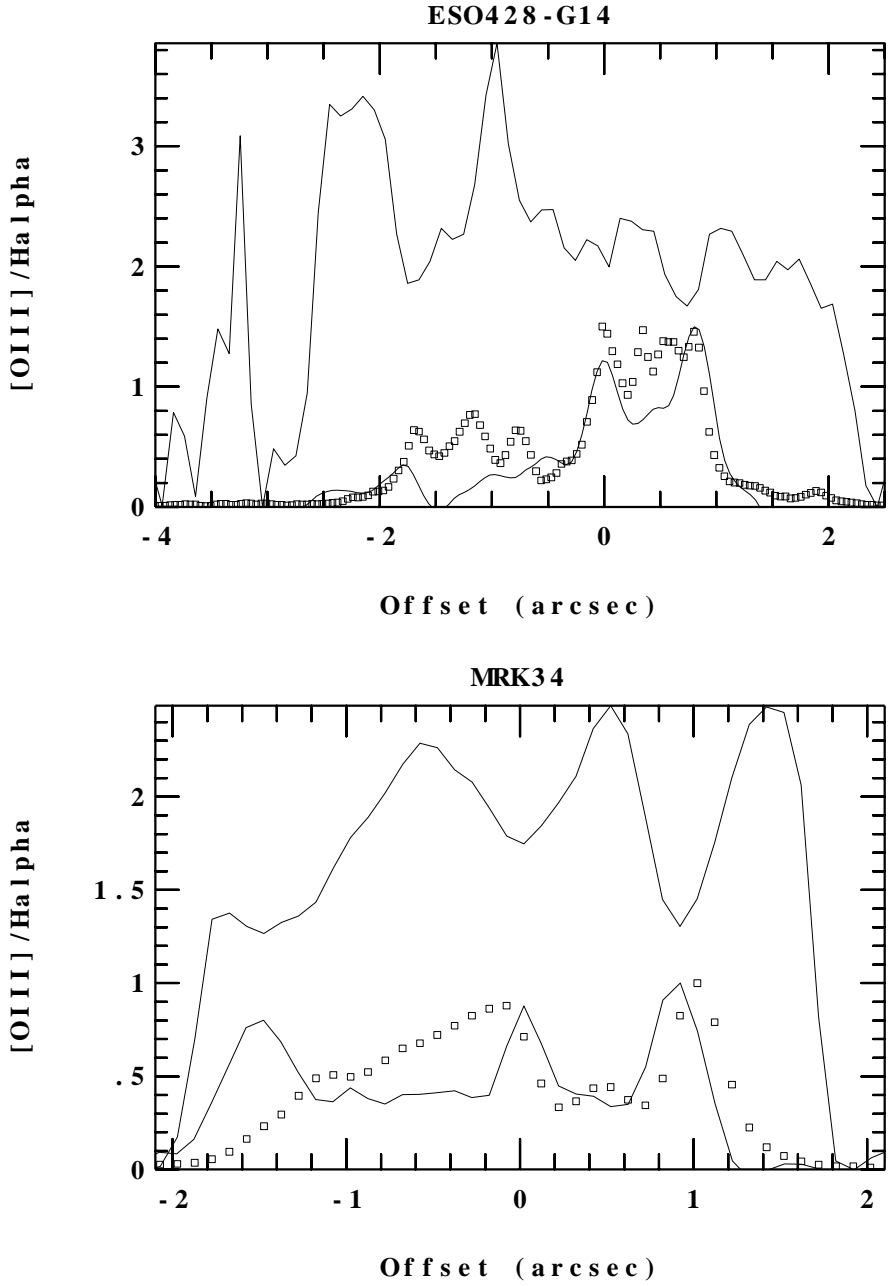}
\caption[]{\label{f10}As Fig.~\ref{f9}. 
Top panel: ESO~428$-$G14 at PA = $-$46$^\circ$ (2 cm map); 
the ordinate scale for
the H$\alpha$+[\ion{N}{2}] brightness is in
units of 
$4.2\times10^{-15}$ erg sec$^{-1}$  cm$^{-2}$ 0\farcs1$^{-2}$.
Bottom panel: Mrk~34 at PA = $-$27$^\circ$ (3.5 cm map); 
the ordinate scale for
the H$\alpha$+[\ion{N}{2}] brightness is in
units of 
$2.0\times10^{-15}$ erg sec$^{-1}$  cm$^{-2}$ 0\farcs1$^{-2}$.
}\end{figure*}

\begin{figure*}
\plotone{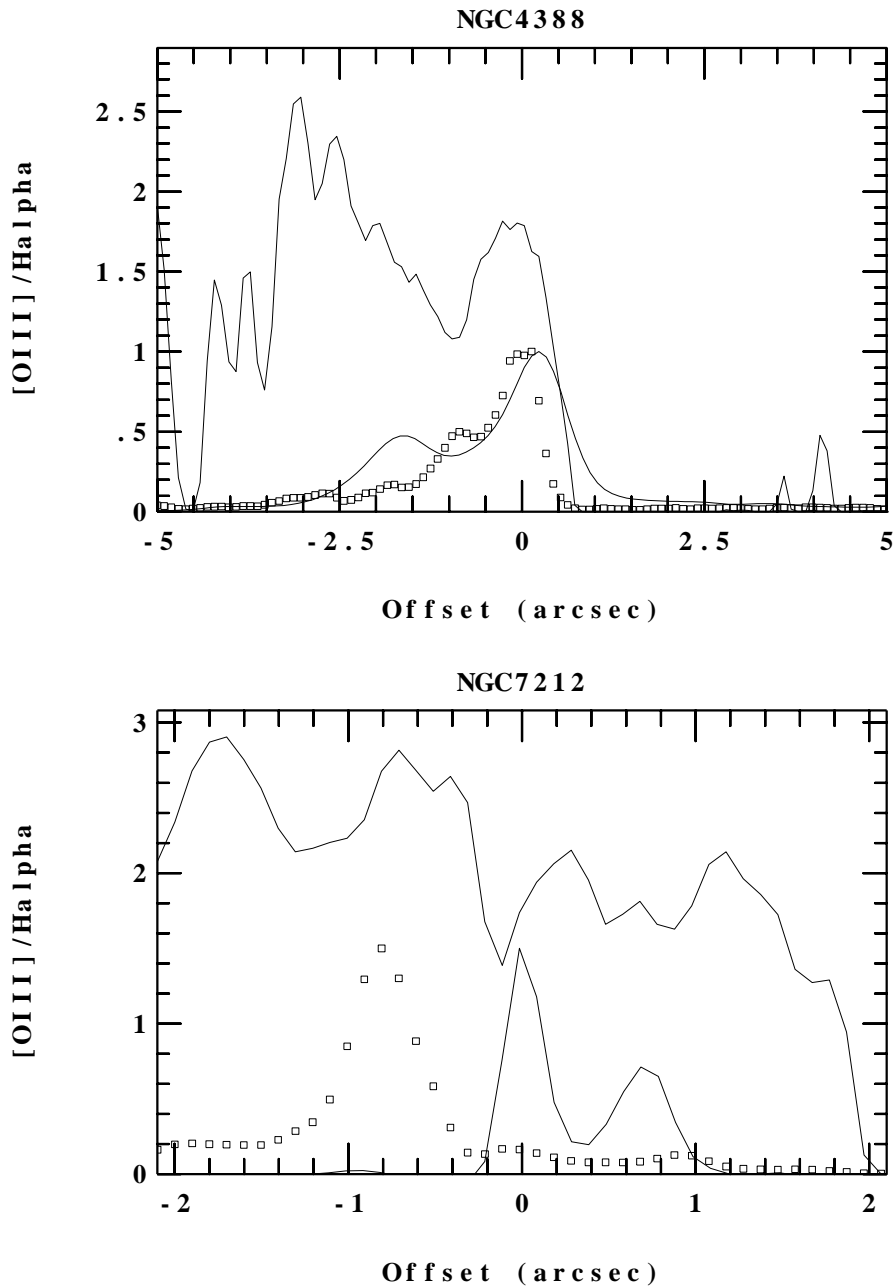}
\caption[]{\label{f11}As Fig.~\ref{f9}. 
Top panel: NGC~4388 at PA = 24$^\circ$ (3.5 cm map); 
the ordinate scale for
the H$\alpha$+[\ion{N}{2}] brightness is in
units of
$2.15\times10^{-15}$ erg
sec$^{-1}$  cm$^{-2}$ 0\farcs1$^{-2}$.
Bottom panel: NGC~7212 at PA = $-$7$^\circ$ (3.5 cm map); 
the ordinate scale for
the H$\alpha$+[\ion{N}{2}] brightness is in
units of
$1.9\times10^{-16}$ erg
sec$^{-1}$  cm$^{-2}$ 0\farcs1$^{-2}$ (Note: this cross section
is significantly offset from the optical nucleus).
}\end{figure*}

Our VLA radio maps are presented in
Figs.~\ref{f6}-\ref{f8}. Cross-sections along the radio axis of our
excitation maps, the H$\alpha$+[\ion{N}{2}] images and the radio maps
for each source are given in Figs.~\ref{f9}-\ref{f11} and overlays of radio and
emission-line maps are shown in Figures~\ref{f11} \&
\ref{f13}. In the following we will discuss each of the galaxies
individually. At the beginning of each paragraph we list the
heliocentric redshift, the distance derived by converting the redshift
to the Galactic Standard of Rest (RC3; deVaucouleurs et al. 1991, and
NED) and using Hubble's law (except for NGC~4388), and the linear
scale corresponding to 0\farcs1---the typical pixel size.

\subsection{Comments on individual galaxies}\label{comments}

\subsubsection{Mrk~348}
\begin{figure*}
\plotone{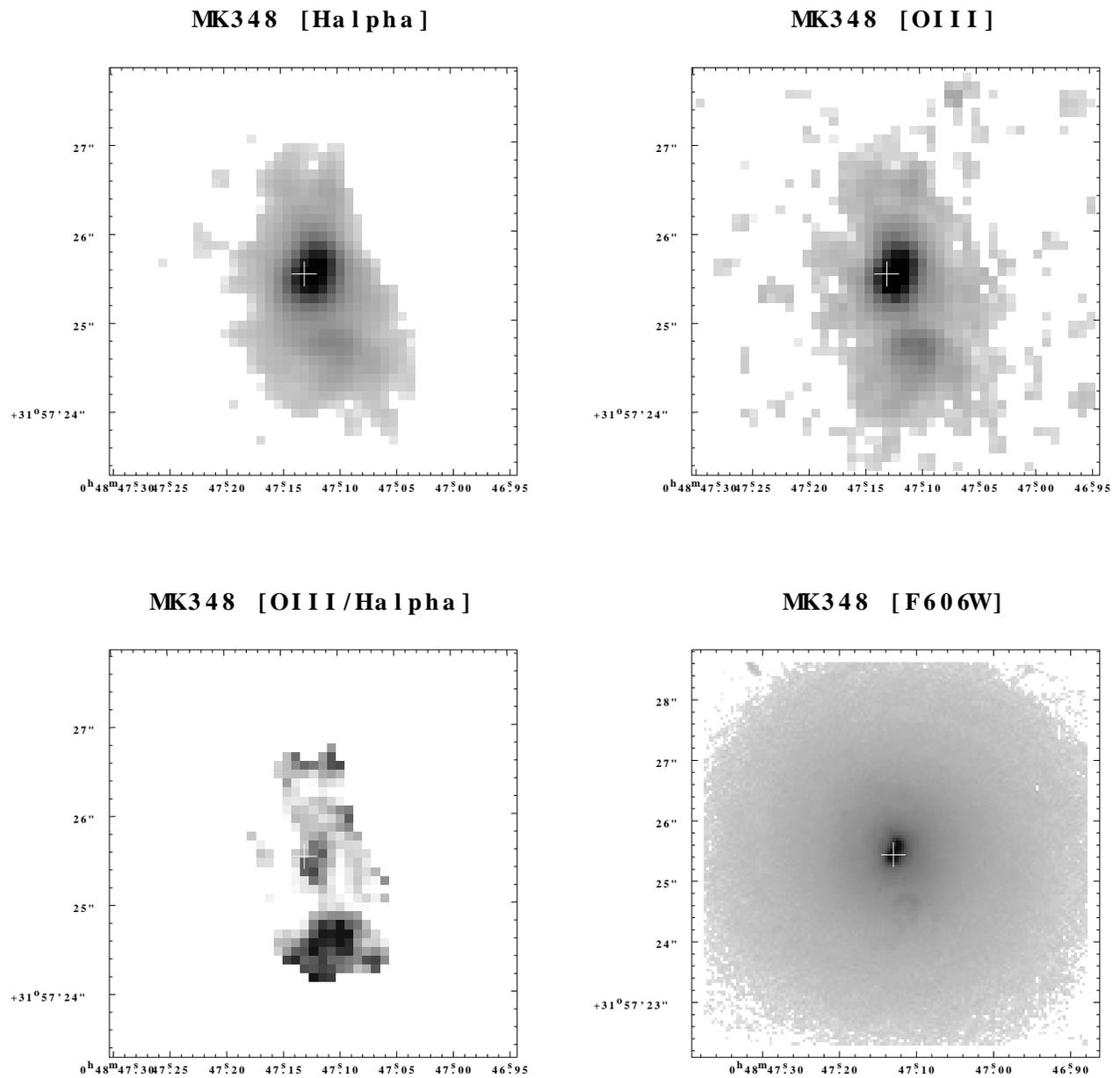}
\caption[]{\label{f1}Mosaic of HST images for Mrk~348 with J2000 astrometric
coordinates. North is up, east is to the left. Upper left: H$\alpha$+[\ion{N}{2}]; upper right: [\ion{O}{3}]; lower left: excitation map,
darker shades correspond to higher
[\ion{O}{3}]-to-H$\alpha$+[\ion{N}{2}] ratios; lower right: broad band
continuum (F606W). The white cross marks the red continuum peak.
}\end{figure*}

Mrk~348 (z=0.01514, D=62 Mpc, 0\farcs1$\equiv$31 pc) is a Seyfert 2
galaxy with a broad H$\alpha$ emission-line in polarized light (Miller
\& Goodrich 1990, Tran 1995). Mulchaey, Wilson, \& Tsvetanov (1997)
and Simpson et al.~(1996) found a possible `ionization cone' in this
galaxy and Neff \& de Bruyn (1983) discovered a triple radio source
along PA 168\arcdeg{} with an extent of 0\farcs2. This direction is
approximately perpendicular to that of the optical polarization (PA
65$^\circ$; Tran 1995). Capetti et al. (1996) observed this galaxy
with the Faint-Object-Camera (FOC) on-board the HST and resolved the
[\ion{O}{3}] emission-line gas in the inner arcsecond into a 0\farcs35
long feature at PA 155\arcdeg. This feature can also be seen as an
elongated nucleus in our lower resolution emission-line images and in
the F606W broad-band filter, which includes H$\alpha$ and
[\ion{N}{2}]$\lambda\lambda$6548,6583 (Fig.~\ref{f1}). A similar, but
less pronounced, elongation is indicated in both line-free continuum
images in the inner 0\farcs3. Figure \ref{f1}, Fig.~\ref{f9}, and
Table~4 show that the line emission is very concentrated within the central
arcsecond. The excitation structure of the most extended gas of the
NLR is difficult to interpret. The southern part could be described as
triangular or fan shaped, typical of `ionization cones' with an
opening angle of $\sim90$\arcdeg. The structure is produced by the
blob of highly excited gas 1\arcsec\ south of the nucleus, also found
by Capetti et al.~(1996). Such a small scale cone would be consistent
with the southern ionization cone claimed by Mulchaey et al.~(1997)
and Simpson et al.~(1996) on larger scales, but on the other hand it
seems that the apex of this cone does not coincide with the optical
nucleus. The good agreement between the VLBI radio and FOC optical
structure (Capetti et al. 1996) suggests that the nuclear engine is
indeed located near the optical nucleus and in that case one would
expect it to coincide with the apex of an ionization cone, even though
the nominal position of the radio source given in Table~3 is, in fact
very close to the apex of the cone.  Another problem, besides the
displacement of the cone's apex from the nucleus, is that the northern
part of the emission-line gas does not exhibit anything resembling a
fan-shape. That could, however, be a result of obscuration or an
inhomogeneous gas distribution.

\subsubsection{Mrk~573}

\begin{figure*}
\plotone{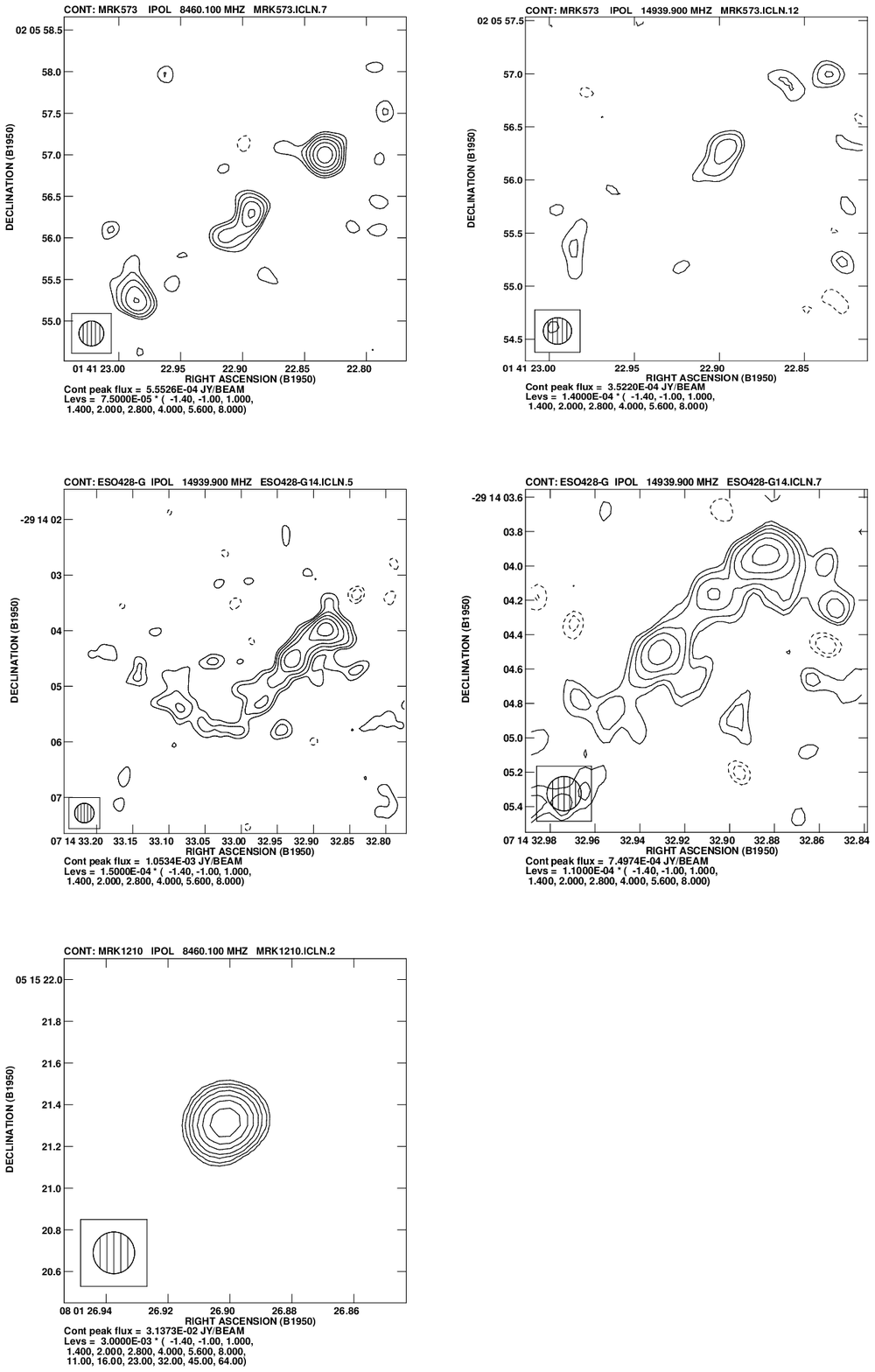}
\caption[]{\label{f6}\label{f7}
VLA maps with B1950 coordinates. North is up, east is to the left.
Upper left: Mrk~573 at 3.5 cm with 0\farcs35 beam.
The peak flux is 0.55 mJy (beam area)$^{-1}$ and contours
are plotted at -1.4, -1 (dotted), 1, 1.4, 2, 2.8, 4, and
5.6 times 0.075 mJy (beam area)$^{-1}$.
Upper right: Mrk~573 at 2 cm with
0\farcs25 beam.
The peak flux is 0.35 mJy (beam area)$^{-1}$ and contours
are plotted at -1.4, -1 (dotted), 1, 1.4, and 2 times 0.14
mJy (beam area)$^{-1}$.
Middle left: ESO~428$-$G14 at 2 cm,
tapered map with 0\farcs35 beam.
The peak flux is 1.05 mJy (beam area)$^{-1}$ and contours
are plotted at -1.4, -1 (dotted), 1, 1.4, 2, 2.8, 4, and
5.6 times 0.15 mJy (beam area)$^{-1}$.
Middle right: enlargement of
ESO~428$-$G14 at 2 cm with 0\farcs2 beam. 
The peak flux is 0.75 mJy (beam area)$^{-1}$ and contours
are plotted at -1.4, -1 (dotted), 1, 1.4, 2, 2.8, 4, and
5.6 times 0.11 mJy (beam area)$^{-1}$.
Lower left: Mrk~1210 at 3.5 cm
with a 0\farcs2 beam. 
The peak flux is 31.4 mJy (beam area)$^{-1}$ and contours
are plotted at  -1.4, -1 (dotted), 1, 1.4, 2, 2.8, 4, 5.6 and 8.0
times 3 mJy (beam area)$^{-1}$.
}\end{figure*}

Mrk~573 (z=0.0173, D$=70$ Mpc, 0\farcs1$\equiv$34 pc) is known from
ground-based observations to have an extended emission-line region
(e.g.~Unger et al. 1987; Haniff, Wilson, \& Ward 1988; Tsvetanov \&
Walsh 1992) with a triple radio source (Ulvestad \& Wilson
1984). Whittle et al.~(1988) found kinematic substructure in the
[\ion{O}{3}] emission-line at the position of the radio hotspots. High
resolution imaging of the NLR of Mrk~573 (Pogge \& De Robertis 1995;
Capetti et al.~1996) reveals a bi-conical excitation structure
enveloping two sets of bow-shock-like emission-line strands outside
the triple radio source (see Fig.~\ref{f12}). The emission-line structure
will be discussed in more detail in Wilson et al.~(1997). Here we
present improved radio maps of this galaxy at 2 and 3.5 cm
(Fig.~\ref{f6}). The 2 cm observations show structure in the three radio
components for the first time. The central radio blob is elongated at
PA$\sim$140\arcdeg{} and has a shape similar to the emission-line
structure associated with the optical peak.

This prompted us to align the peak in the radio emission with the
optical continuum peak (Fig.~\ref{f12}). Pogge (1996) and Capetti et
al.~(1996), however, have argued that the real optical nucleus is
hidden behind a dust lane, 0\farcs3 to the NW. Both registrations are,
to within the errors, compatible with the astrometric positions, so we
cannot decide between them.

In the 3.5 cm map the SE radio component is extended almost
perpendicular to the source axis (Fig.~\ref{f6}), which may indicate that the
plasma flow is redirected inside the bow-shocks, as is also suggested
by the 2 cm map. Comparison of the emission-line and the radio
structure shows that the radio hotspots are located on the tips of two
emission-line strands coming out of the nucleus, just inside the inner
two emission-line arcs. In tapered, low-resolution 3.5 cm maps, we
also find weak radio emission connecting the three blobs and some very
faint radio emission may be associated with the outer two
emission-line arcs, some 2\farcs9 to the SE and 3\farcs6 to the NW of
the nucleus. When comparing our highest resolution 3.5 cm map, with a
tapered 2 cm map we find that the hotspots have steep radio spectra
with $\alpha$ ($S_\nu\propto\nu^\alpha$) between $-0.7$ and $-1$ for
the NW hotspot and $\alpha\sim-0.5$ for the SE hotspot. The central
blob, however, has an inverted or flat spectrum
($\alpha\sim+0.2\pm0.2$)---another clear indication that this is the
radio nucleus. Figure~\ref{f9} (bottom panel) also shows that the radio
hotspots coincide with regions of reduced excitation. 

\subsubsection{ESO~428$-$G14}
\begin{figure*}
\centerline{\fbox{Color Plate 1 (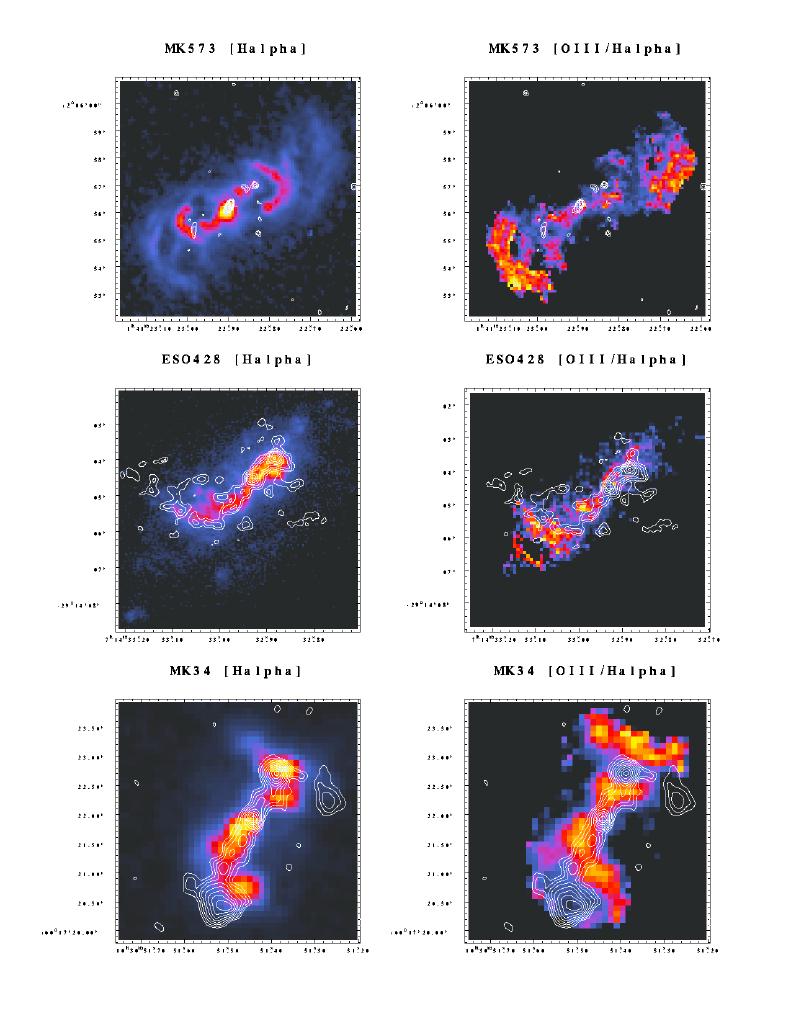)}}
\caption[]{\label{f12}H$\alpha$+[\ion{N}{2}] images and
[\ion{O}{3}]/(H$\alpha$+[\ion{N}{2}]) excitation maps with radio
contours overlaid. The coordinates are the B1950 coordinates of the
radio maps. North is up, east is to the left. 
Upper left:  Mrk~573, H$\alpha$+[\ion{N}{2}] and 2 cm; 
upper right:  Mrk~573, [\ion{O}{3}]/(H$\alpha$+[\ion{N}{2}]) and 2 cm;
middle left: ESO~428$-$G14, H$\alpha$+[\ion{N}{2}] and 2 cm; 
middle right: ESO~428$-$G14, [\ion{O}{3}]/(H$\alpha$+[\ion{N}{2}]) and 2 cm;
lower left:  Mrk~34, H$\alpha$+[\ion{N}{2}] and 3.5 cm; 
lower right:  Mrk~34, [\ion{O}{3}]/(H$\alpha$+[\ion{N}{2}]) and 3.5 cm.
}\end{figure*}

\begin{figure*}
\centerline{\fbox{Color Plate 2 (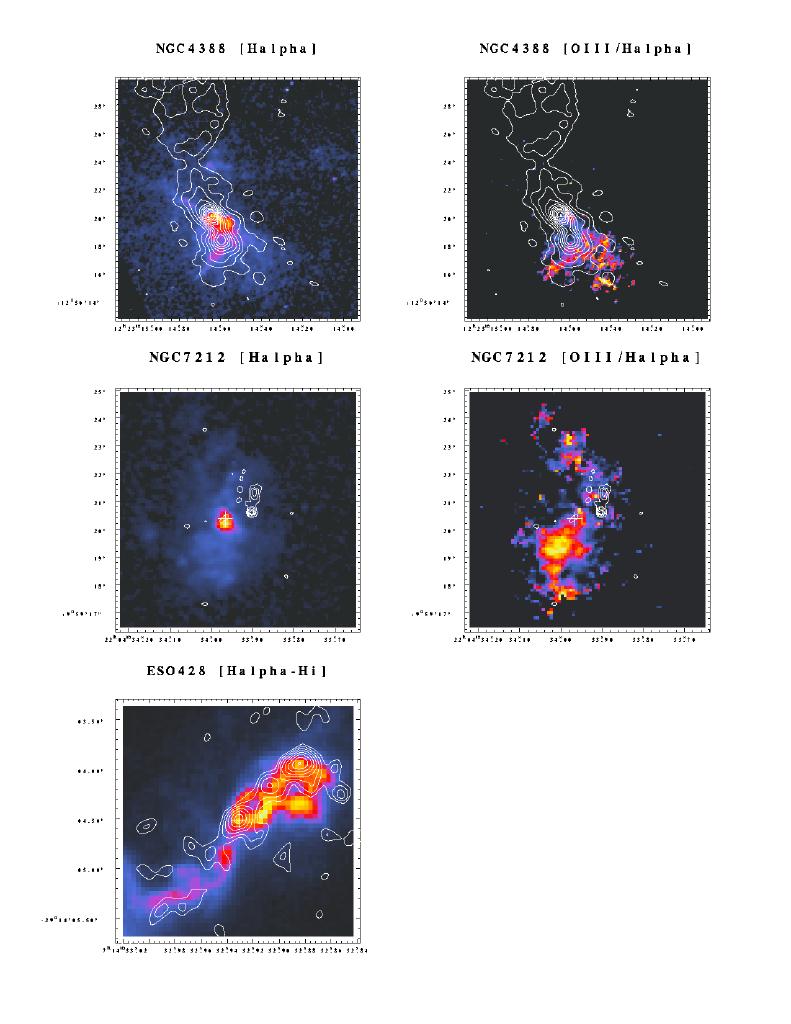)}}
\caption[]{\label{f13}As Fig.~\ref{f12}.
Upper left: NGC~4388, H$\alpha$+[\ion{N}{2}] and 3.5 cm; upper right:
NGC~4388, [\ion{O}{3}]/(H$\alpha$+[\ion{N}{2}]) and 3.5 cm; middle
left: NGC~7212, H$\alpha$+[\ion{N}{2}] and 3.5 cm; middle right:
NGC~7212, [\ion{O}{3}]/(H$\alpha$+[\ion{N}{2}]) and 3.5 cm; lower
left: enlargement of the figure `eight' distribution of
H$\alpha$+[\ion{N}{2}] in ESO~428$-$G14 with the high-resolution 2 cm
map overlaid. The radio coordinates for NGC~4388 are shifted to the
frame of Carral et al.~(1990).
}\end{figure*}

ESO~428$-$G14 (z=0.00544, D$=19.0$ Mpc, 0\farcs1$\equiv$9.2 pc) was
described in more detail in Falcke et al.~(1996b), where we also
discussed the optical-radio registration. The emission-line images
presented there showed an unusual structure consisting of thin strands
of ionized gas which form a figure `eight' to the NW of the
nucleus. Our new radio maps (Fig.~\ref{f6}) substantially improve on the
earlier radio map by Ulvestad \& Wilson (1989) in both sensitivity and
resolution. The higher resolution 2 cm map shows a narrow jet to the NW
of the nucleus terminating in a hot spot which coincides with the
location where the emission-line strands end (Fig.~\ref{f12}). The radio jet is
unresolved perpendicular to its axis with a beam of 0\farcs2. In
Falcke et al.~(1996b) we speculated that the helical emission-line
strands would wrap around a collimated radio-jet, with the strands
formed in the boundary layer between jet and ISM. Not only do we find
this jet in the radio now, but the images also show that the maximum
separation of the optical strands transverse to the jet axis
($\sim$0\farcs3, Fig.~\ref{f12}) is wider than the radio-jet itself, lending
support to the boundary layer hypothesis. One problem, however, is the
exact radio/optical registration for ESO~428$-$G14, since with the
current alignment, the radio jet does not go right through the middle
of the figure `eight' (see enlargement Fig.~\ref{f13}). A shift of the radio
by 0\farcs15 to the south would solve this problem, but would also
imply an offset between radio and optical nuclei. Until high
precision astrometry is available for this galaxy, the exact relation
between radio and optical emission will remain somewhat uncertain.

Nevertheless, a few important observations can be made. At the NW tip
(0\farcs85 from the nucleus) of the jet the radio emission is
associated with the northern optical strand (Fig.~\ref{f13}) and the NW
radio hot spot is associated with a region of low emission-line
excitation (Figs.~\ref{f10} \& \ref{f12}). Near the NW radio hotspot,
both the radio jet and the optical strand bend towards the south for
0\farcs3 until they intersect the southern strand.  Further to the NW,
$\sim$1\farcs2 from the nucleus, we also find some radio emission
which might be associated with the faint linear, high-excitation
[\ion{O}{3}] feature which sticks out of the figure `eight' towards
the NW.

The SE part of the radio jet has a lower surface brightness than the
NW structure and consists of at least one faint strand roughly
following, but not coincident with, the optical strands (Figs.~\ref{f6} and
\ref{f12}). There is some indication, that, roughly at the same distance from
the nucleus as the NW hotspots, the SE radio jet splits into two, with
the northern jet being much fainter than the southern one. This
northern jet is, in fact, more clearly visible on our 3.5 cm map (not shown
here) and appears to be significant. Interestingly, the location where
the jet splits coincides with the region of highest excitation in the
ionized gas (Fig.~\ref{f12}). Towards the SE end, the radio jet clearly bends
towards the north, as does the high-excitation ionized gas.

\subsubsection{Mrk~1210}
\begin{figure*}
\plotone{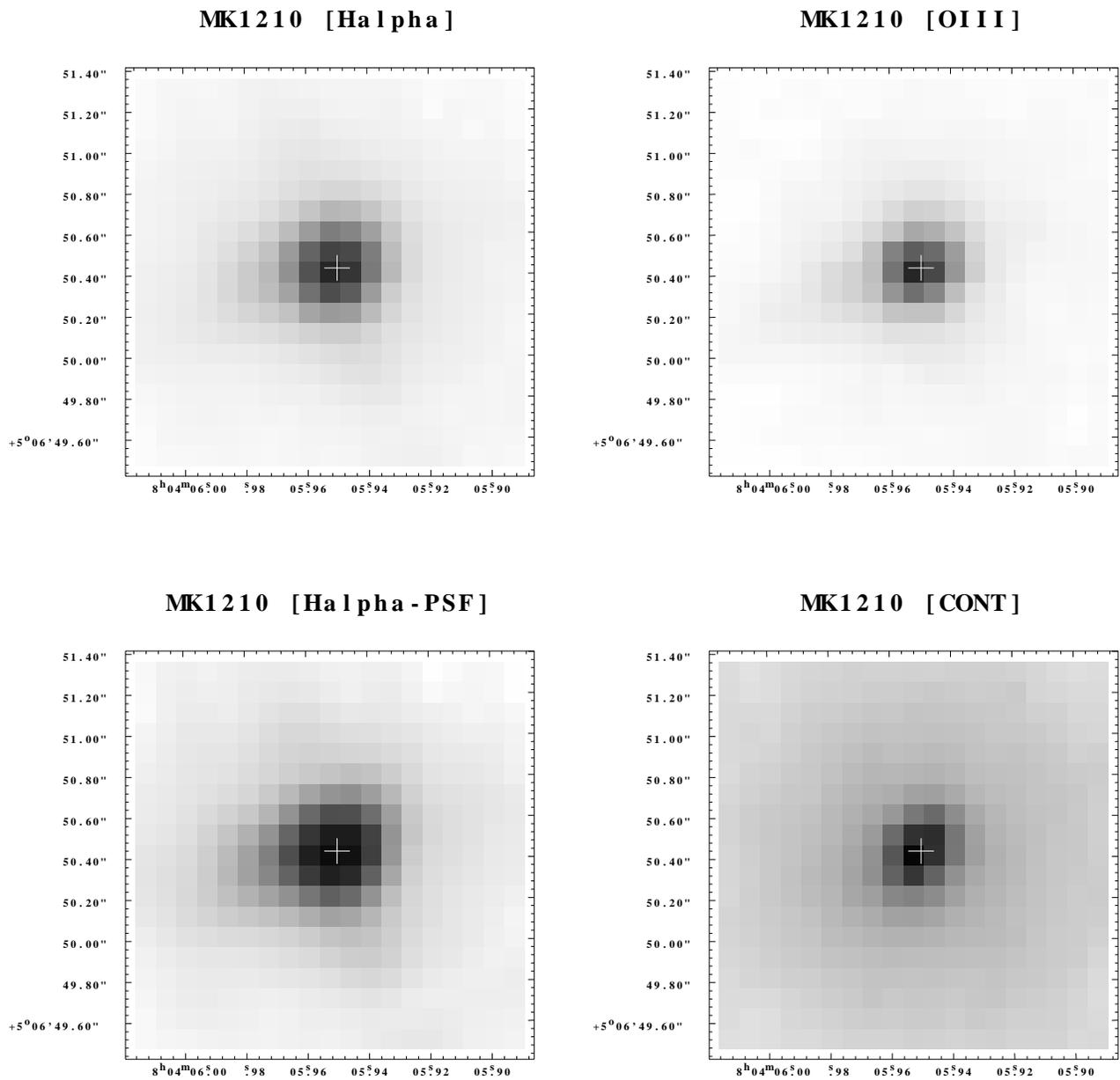}
\caption[]{\label{f2}Mosaic of HST images for Mrk~1210 with J2000 
coordinates. North is up, east is to the left. Upper left: H$\alpha$+[\ion{N}{2}]; upper right: [\ion{O}{3}]; lower left: PSF subtracted
H$\alpha$+[\ion{N}{2}] image; lower right: red continuum image. The white
cross marks the red continuum peak.
}\end{figure*}

Mrk~1210 (z=0.0135, D$=52$ Mpc, 0\farcs1$\equiv$25 pc) is another
Seyfert 2 galaxy selected because of its broad emission-lines seen in
polarized light (e.g.~Tran 1995). The emission-line structure we find
in this galaxy is the most compact in our whole sample. The emission
is dominated by a single central peak (Fig.~\ref{f2}). We
have compared radial profiles of the nuclear emission-line peak with
those of stars in our FOV and could not find any significant
differences. Hence, this compact, bright part of the nuclear line
emission is $<$40pc in extent. 

We then tried to subtract the central point source to look for any
faint extended emission. This is a difficult task, since the
point-spread function (PSF) for the LRFs is a function of wavelength
and position on the four chips, and will be different for each
exposure. Unless one explicitly makes a separate PSF observation it is
almost impossible to find an exactly matching PSF in the archive.
Hence, we composed an average PSF from 5 stars which were visible in
the FOV of our H$\alpha$+[\ion{N}{2}] image of Mrk~1210 and which had
more than 1000 counts. The stars were shifted onto a common center,
scaled and weighted according to their peak flux, and then
co-added. Finally, we subtracted this average PSF image from the
continuum-subtracted, but not rotated, H$\alpha$+[\ion{N}{2}] image of
Mrk~1210 until the central pixel had the average flux value of its
neighbors. The result is shown in Fig.~\ref{f2}, in which the PSF subtracted
image has been rotated to the cardinal orientation. A PSF subtraction
of the [\ion{O}{3}] was not possible since the FOV of this image did
not include any stars with enough counts to determine the PSF
reliably.

After subtraction the central point source, we find some faint,
extended H$\alpha$+[\ion{N}{2}] emission with a diameter of
$\sim0\farcs5$ (Fig.~\ref{f2}). Some of the structure visible may, however,
still reflect uncertainties in the PSF.  If we divide the [\ion{O}{3}]
by the H$\alpha$+[\ion{N}{2}] image (without PSF subtraction), we find, an unusually
high [\ion{O}{3}]-to-H$\alpha$+[\ion{N}{2}] ratio of 5 for the central peak while
the faint extended emission has [\ion{O}{3}]/(H$\alpha$+[\ion{N}{2}])$\sim2$
(determined in the region not dominated by the instrumental `spider'
structure of the central peak).

Not surprisingly, the radio structure is also very compact, but is
slightly resolved (Fig.~\ref{f7}). We find a total, integrated
flux of 39 mJy at 3.5 cm (31.4 mJy peak), which is consistent with
earlier observations by Antonucci \& Barvainis (priv. comm.) who
obtained 39 mJy (total) at 3.5 cm and 16.3 mJy (total and peak) at
K-band, thus yielding a spectral index of $\alpha=-0.7$ for the
peak. This indicates a compact, non-variable steep-spectrum
core. Parkes-Tidbinbilla interferometer (PTI) observations at 13 cm by
Norris et al.~(1990) found only 48 mJy, much less than the peak flux
expected from extrapolation of the higher frequency observations ($\sim$80
mJy). This difference could be due to either a low-frequency turnover,
e.g. caused by free-free absorption, or resolution of the
steep-spectrum core, even though the resolution of the
PTI at 13 cm ($\sim$0\farcs1) is not  much higher than that of the VLA
at 3.5 cm ($\sim$0\farcs2).

Finally, we note that if the slight elongation to the SE visible in
the PSF subtracted H$\alpha$+[\ion{N}{2}] image (but see remarks
above) and possibly also in the radio map is real, then this direction
would be roughly perpendicular to the optical polarization vector
(PA=29$^\circ$) as measured by Tran (1995). Only high-resolution radio 
mapping of this source, e.g. with the VLBA at lower frequencies, may
be able to resolve this question and show whether there is an
underlying jet-structure in this source.

\subsubsection{Mrk~34}
\begin{figure*}
\plotone{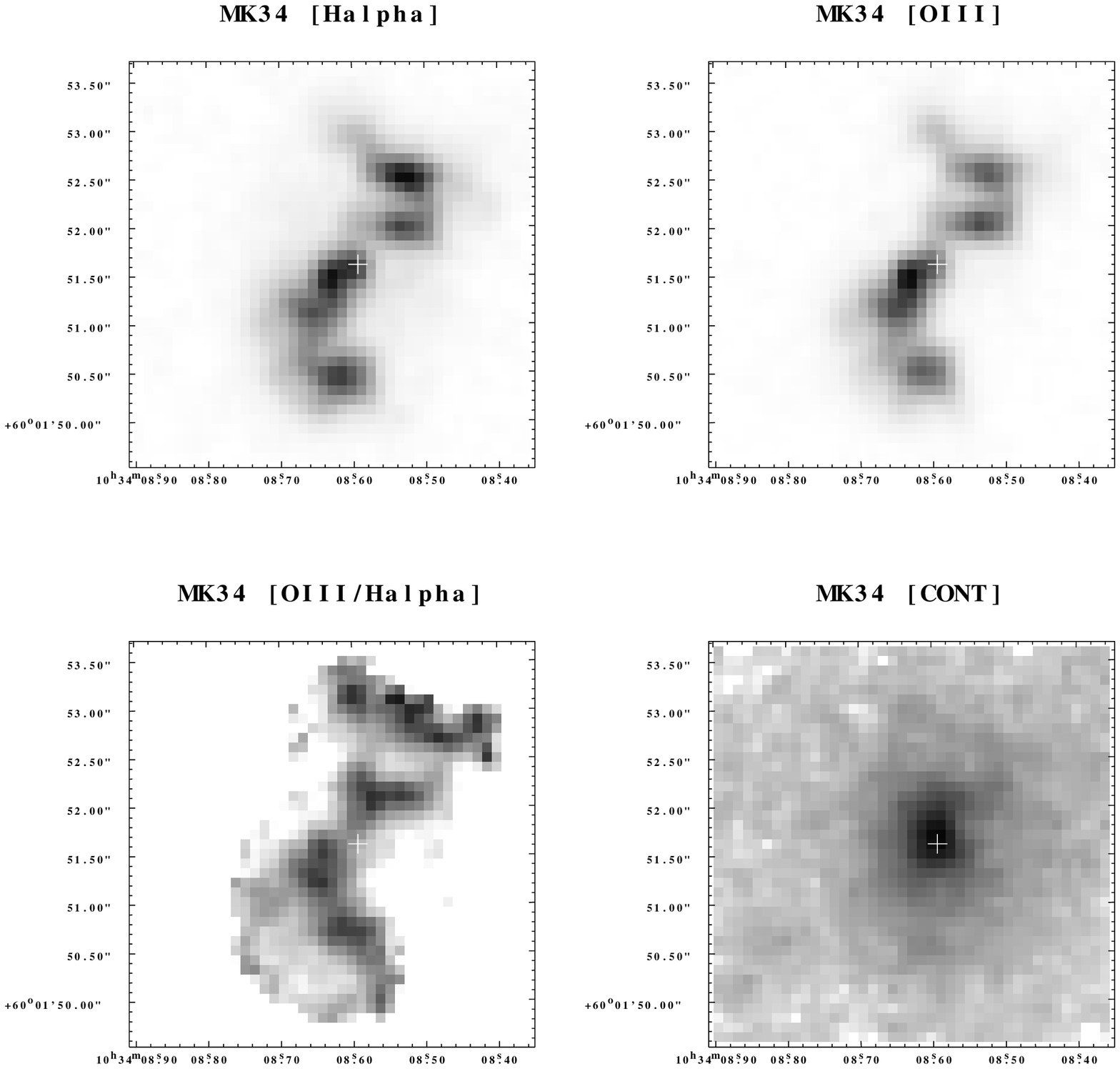}
\caption[]{\label{f3}Mosaic of HST images for Mrk~34 with J2000 astrometric
coordinates. North is up, east is to the left. Upper left: H$\alpha$+[\ion{N}{2}]; upper right: [\ion{O}{3}]; lower left: excitation map,
darker shades correspond to higher
[\ion{O}{3}]-to-H$\alpha$+[\ion{N}{2}] ratios; lower right: red
continuum image. The white cross marks the red continuum peak.
}\end{figure*}

Mrk~34 (z=0.0505, D$=205$ Mpc, 0\farcs1=100 pc) is the most distant
and most luminous galaxy in our sample. Its [\ion{O}{3}] luminosity of
$3\times10^{42}$ erg sec$^{-1}$ is comparable to that of an average
radio-quiet, low-redshift ($z<0.5$) quasar in the Palomar-Green quasar
sample (Boroson \& Green 1992). From ground-based optical (Haniff et
al.~1988) and VLA radio observations (Ulvestad \& Wilson 1984), it is
known that this galaxy has an extended structure with two
emission-line and radio lobes. Our emission-line images (Fig.~\ref{f3}) show
an S-shaped structure resembling a seahorse. The inner part consists
of one emission-line strand which seems to wiggle around a central
axis along PA$\simeq-37$\arcdeg{} until it terminates and bends
through $\sim$90\arcdeg{} at both ends. In the excitation map (Fig.~\ref{f3})
we find a bi-polar high-excitation morphology, but the structures on
opposite sides of the nucleus are not anti-parallel and the optical
nucleus is offset from the apices of the structures.  Quite obviously
the optical strand associated with the jet leaves its imprint in
the excited gas, yielding an asymmetric appearence in the south.

Our radio maps also improve substantially upon earlier images. The
maps show a radio jet with hotspots at each end (Figs.~\ref{f8} and
\ref{f12}). The overall morphology is very straight along
PA=$-$27\arcdeg{} with some small wiggles in the jet. This structure
is unlike radio galaxies, since Mrk~34 not only has two pronounced
hotspots, like luminous FR\,II radio galaxies, but also well defined,
two-sided jets, as in low-luminosity FR\,I radio galaxies.  The
spectral indices we find for the jet from our X- and C-band data are
extremely steep. The nuclear region has $\alpha\simeq-1$ and the NW
hotspot $\alpha\simeq-1.3$, while the spectrum of the SE hotspot is
even steeper with $\alpha\simeq-1.7$. However, these numbers should be
treated with caution, as resolution effects tend to steepen spectra
taken in a single VLA configuration.

Since the astrometric positions (Table~3) indicate that the brightest
peak in the jet is very close to the optical nucleus, we have aligned
those two peaks.  The emission-line structure then aligns very closely
with the radio structure, so the ionized gas is physically associated
with the radio jet (Fig.~\ref{f12}). Remarkably, the two radio
hotspots coincide with pronounced regions of low excitation
(Figs.~\ref{f10} and \ref{f12}; see also Fig.~\ref{f3}), possibly
because the ionized gas is compressed to a higher density, so that the
ionization parameter is reduced. Whittle et al.~(1988) found that the
gas associated with the radio hot spots has kinematically distinct
properties and is clearly separated in velocity by several hundred
km/sec from the ambient gas.

\subsubsection{NGC~4388}
NGC~4388 (z=0.00842, D$=16$ Mpc, 0\farcs1=7.8 pc) is assumed to be at
the core of the Virgo cluster, for which we adopt a distance of 16 Mpc
(van den Bergh 1996), and thus this galaxy is the closest in our
sample. It is know to have extended ionization cones from
ground-based observations (Pogge 1988; Corbin, Baldwin, \& Wilson
1988).

The host galaxy is an edge on spiral and hence dust obscuration plays
an important role in the interpretation of the optical structure. Our
low signal-to-noise continuum image (Fig.~\ref{f4}) shows two wedges
of reduced emission, demonstrating the presence of obscuring dust
bands along the disk of its galaxy right into the nuclear region.
Several earlier papers have already suggested that NGC~4388 is an
obscured Seyfert~1 galaxy on the basis of weak, broad H$\alpha$ lines
(Filippenko \& Sargent 1985, see also Shields \& Filippenko 1996), an
inferred, hidden ionizing continuum source (e.g.~Colina 1992), and the
presence of hard X-rays (Hanson et al. 1990).  The velocity and
excitation structure of the NLR is very complex (Corbin et al.~1988;
Veilleux 1991) and might be described by rotation plus outflow.  VLA
maps have shown a central double source and a plume of radio plasma to
the north of the optical nucleus (Stone, Wilson, Ward 1988; Hummel \&
Saikia 1991).

Our H$\alpha$+[\ion{N}{2}] image (Fig.~\ref{f4}) shows ionized gas extending
$\sim$6\arcsec{} to the north and the south of the nucleus. Both the
[\ion{O}{3}] image and the excitation map (Fig.~\ref{f4}) show a bright
region of high-excitation gas to the south, including a very well
defined ionization cone, which is the exact continuation to smaller
scales of the ionization cone found from ground-based images
(e.g.~Pogge 1988). A very interesting feature is a bright spike in the
emission-line images, 2\farcs5 south of the optical nucleus, that runs
at a PA of 140\arcdeg{} through the middle of the ionization cone.

Our 3.5 cm map of this galaxy (Fig.~\ref{f8}) confirms the structure found in
earlier VLA observations but shows more detail. The plume to the north
is very pronounced and opens up dramatically after the radio outflow
is tightly constricted (4\arcsec\ to the north of the radio nucleus),
resembling one half of a limb-brightened hour glass. The northern blob
of the compact double source near the center of the galaxy appears to
be the nucleus, since it has a flat spectrum down to at least 2 cm (see
Carral, Turner, \& Ho 1990).

As discussed in Section~\ref{registration}, the exact alignment of
radio and optical images of NGC~4388 requires care. Previous
ground-based measurements of the optical nucleus (AE90) give a
position almost 3\arcsec{} south of the radio nucleus. This is,
however, an artifact resulting from the highly asymmetric
emission-line region (Fig.~\ref{f4}) and the smoothing effect of atmospheric
seeing. Our new registration (Fig.~\ref{f13}), using the nearby field star
measured by Argyle (see Sec.~2.3) and the radio position of the
nucleus from Carral et al~(1990), shows that, in fact, the optical
nucleus as seen with HST is within 0\farcs45 of the radio nucleus and
the apex of the ionization cone. The radio blob 2\arcsec\ to the south
of the nucleus is clearly associated with the spike seen in the
emission-line images and this relationship is strongly suggestive of
an interaction between the radio jet and a dense cloud in the
ISM---faint radio emission also extends along this spike. We also note
from the lowest radio contours that there seems to be low-surface
brightness radio emission filling the whole inner part of the
ionization cone.

\subsubsection{NGC~7212}
NGC~7212 (z=0.0266, D$=110$ Mpc, 0\farcs1=53 pc) is a Seyfert 2 galaxy
in an interacting system of galaxies\footnote{Wasilewski classified
this as a triple system.  The component 40\arcsec{} SW of the
emission-line nucleus is not visible in the image shown here. On the
archival F606W image this component is much smaller than the two
merging galaxies and highly elongated. In addition the large FOV of
the F606W image shows the rich structure and tidal tails created by
the debris of this merger.} (Wasilewski 1981).  Both galaxies in the pair
visible in our continuum image (Fig.~\ref{f5}) exhibit multiple dust
lanes. The NLR structure is elongated and diffuse with a strong
central peak near the continuum peak and an overall structure
resembling Mrk~348. The excitation map is more difficult to interpret
since there is no clear ionization cone visible, and the emission-line
distribution in the north is partly obscured by the dust lanes. It
might be said that the distribution of excitation to the south is
consistent with a ragged ionization cone.

The VLA map of NGC~7212 (Fig.~\ref{f8}) shows a compact, double source
separated by 0\farcs7 in the north-south direction, with the northern
blob being slightly elongated. The small angular extent in the radio
is another similarity to Mrk~348.

Using the astrometry based upon Argyle's star in the FOV of this
galaxy, we find that the compact, southern radio blob, is offset by
almost 1\arcsec{} to the NW of the optical nucleus
(Fig.~\ref{f13}). This is substantially larger than the typical errors
in these astrometric observations and either indicates a not yet
understood systematic error or is indeed significant. Our new position
for the optical nucleus deviates a few tenth of an arcsecond from the
position given in Clements (1983), but the basic result remains the
same with both positions, suggesting that the offset is real. Hence,
this offset might indicate the presence of a strong dust lane in the
nuclear region. Given our new position the southern lobe of NGC~7212
seems to coincide with a low-excitation region (Fig.~\ref{f13}), an
effect we have already seen in some of the other galaxies in our
sample.

In general, we confirm the results of Tran (1995), who suggested that
dust obscuration plays a significant role in this source (see also the
broad-band continuum image in Fig.~\ref{f5}) and found a jet extending
up to 10\arcsec\ from the nucleus at PA=$-10^\circ$ in ground-based
[\ion{O}{3}] and H$\alpha$ images---exactly the same direction as the
axis of the newly found double radio source on a much smaller spatial
scale. As noted by Tran (1995), this direction is essentially
perpendicular to that of the optical polarization (PA 93$^{\circ}$),
suggesting that the obscuring torus plus scattering region model
may apply to NGC~7212.

\section{Discussion} 
We have observed a sample of seven Seyfert 2 galaxies with HST in the
H$\alpha$+[\ion{N}{2}] and [\ion{O}{3}] emission-lines and with the
VLA at 3.5 cm and 2 cm to study the morphology of the NLR and its
interaction with the radio outflows. We were able to resolve the NLR
in six of the seven galaxies and in all cases found a very elongated
emission-line region. In the four galaxies (Mrk~573, ESO~428$-$G14,
Mrk~34, NGC~4388) whose NLRs have the largest angular extent we find
bi-polar structures in excitation maps, which can be interpreted as
either the `ionization cones' expected in the unified scheme or
widening, self-excited matter outflows.  In all cases where an
`ionization cone' was known from ground-based observations (Mrk~573,
NGC~4388, and possibly Mrk~348), we found an equivalent counterpart on
the sub-arcsecond scale with a similar opening angle. This
continuation and the straightness of the cone edges in NGC~4388 and
Mrk~573 argue in favor of some kind of obscuring `torus' with well
defined inner edges around the nucleus. Nevertheless, we find
substantial structures within those `cones' which indicate that, in
the nuclear region, additional processes (such as jet/ISM
interactions), besides anisotropic radiation, have a major impact on
the excitation of the gas.

Our radio observations have resulted in high quality maps of the
galaxies in our sample, most of which show clear evidence for
radio-jets and/or lobes. Seyfert galaxies show a considerable
diversity in their radio structures with various morphologies: narrow
jets (ESO~428$-$G14), triple structures with a core and two hotspots
(Mrk~573), jets plus two hotspots (Mrk~34), radio plumes and
limb-brightened lobes (NGC~4388), etc. This diversity suggests that
Seyfert jets are much more influenced by jet-ISM interaction than
FR\,I and FR\,II radio galaxies, which may not be surprising since the
radio power of Seyfert galaxies is orders of magnitudes smaller than
in radio galaxies. The existence of this interaction is in fact well
documented by the comparison between the radio continuum and
emission-line images (Figs.~\ref{f12} \& \ref{f13}). In those galaxies
with polarized optical light (e.g.~Tran 1995) we find that the
polarization vector is perpendicular to the axis of the radio
structure and the NLR (Mrk~348, NGC~7212, and possibly also in
Mrk~1210). In several cases we also note that the excitation structure
is influenced by the jet, e.g. radio hotspots seem to coincide with
regions of reduced excitation (Mrk~573, Mrk~34, and probably also
NGC~4388 and NGC~7212), while high-excitation regions are often found
downstream of radio hotspots or regions where the jet bends or comes
to an end (the SE end of ESO~428$-$G14 and the southern hotspot of
NGC~4388). It is quite obvious that there is not only a close
correlation between the radio and the emission-line morphologies, but
that {\it the radio jet-ISM interaction is an important effect which
strongly influences the excitation and morphology of the NLR}.

For example, the emission-line arcs in Mrk~573 wrap around the radio
lobes. Our 2 cm maps (Figs.~\ref{f6} and \ref{f12}) also shows that one of the
radio lobes is elongated and bends in the region where the interaction
is thought to occur. Further, the reduced excitation (i.e.~
[\ion{O}{3}]/(H$\alpha$+[\ion{N}{2}]) ratio) of the inner
emission-line arcs is consistent with a lower ionization parameter 
which can be understood if the arcs represent gas which has passed
through a radiative bow shock, cooled and increased in density. A
similar effect is seen in  Mrk~34, in which the radio lobes coincide
with a low-excitation region, while the jet itself is surrounded by
high-excitation gas. The wiggles seen in the radio jet of this galaxy
could possibly be interpreted as some kind of Kelvin-Helmholtz
instabilities between the radio plasma and the surrounding, ionized
gas.

ESO~428$-$G14 exhibits well-collimated, irregular emission-line
strands on one side and a figure ``eight'' morphology on the
other. The latter has been interpreted as two helical emission-line
strands wrapping around a radio jet (Falcke et al.~1996b). Our new,
high-resolution maps indeed show a filamentary radio jet which
supports this idea: the NW part of the radio jet is straight with the
emission-line helix possibly surrounding it, while the SE radio jet
splits into two at $\simeq$1\arcsec (90 pc) from the nucleus, again
with a close relationship to the emission-line strands. However, since
there is no possible registration in which the radio jet and the
emission-line strands coincide precisely (there always seems to be a
slight offset between those two structures), it seems likely that here
too the emission-line gas wraps around the radio-outflow.

Another example of an interaction between a radio jet and ambient gas
is found in NGC~4388, where a bright spike is seen in the ionized gas
at the end of the southern jet, while the radio plasma to the north
flows apparently unhampered out of the galaxy disk and forms a large
($\sim$ 1 kpc) radio plume. Our 3.5 cm map also indicates that this
radio plume may be hollow and hourglass shaped. This structure is
reminiscent of the radio lobe found, for example, in NGC~3079
(Seaquist et al.~1978) and a few other galaxies (Ford et
al. 1986). This kind of limb-brightened radio lobe stands in marked
contrast to the well-collimated, stranded jets in ESO~428$-$G14 and
NGC~4258 (e.g.~Cecil, Wilson, \& Tully 1992). An important difference
is that NGC~3079 and NGC~4388 have jets which escape almost
perpendicular to the galaxy plane, while in NGC~4258 and perhaps
ESO~428--G14 the jet appears to be directed into the disk of the
galaxy. The difference in radio morphology may then be ascribed to the
much higher external gas density when the jet is in the disk than in
the galaxy halo.

Moreover, even the sources with compact NLRs support the jet-ISM
interaction scenario, since all three such sources also have
correspondingly compact radio structures. These sources happen to be
those selected for HST observations on the basis of having hidden
broad line regions, visible in polarized light, rather than having
extended emission-lines in ground-based observations. The most extreme
case of a compact NLR is Mrk~1210, in which the line emission is
dominated by a point source ($<40$pc) and which has the highest
[\ion{O}{3}]/(H$\alpha$+[\ion{N}{2}]) ratio.

Hence, the most important conclusion that comes out of this study is
not only that an expectation of the unified scheme---`ionization
cones'---is tentatively supported by the HST observations, but also
that the influence of radio jets and their angle with respect to the
galaxy plane cannot be ignored when discussing the structure,
kinematics, and excitation of the NLR of Seyferts. Specifically, 
there seems to be a tendency for radio lobes to coincide with lower
excitation regions, presumably a result of compression of the ambient
gas.

As a consequence of this compression one would also expect that
compact regions with high emission-line brightness (e.g the inner arcs
in Mrk 573) have high density and thus low [\ion{O}{3}]/H$\alpha$ +
[\ion{N}{2}] ratio.  Figure \ref{f14} shows diagrams in which we have plotted
the emission line brightness versus the [\ion{O}{3}]/H$\alpha$ +
[\ion{N}{2}] ratio for each pixel in the four images with the largest
number of usable pixels (i.e. with emission-line brightness above the
3$\sigma$ noise level). The upper envelope for all of these graphs has
a characteristic peaked shape, indicating that the brightest features
have only moderate excitation. In Mrk~573 and NGC~4388, the highest
excitation is clearly found in the lowest brightness regions, as
expected if variations in both of these parameters are primarily
determined by variations in gas density. However, no such clear
relationship is seen for ESO~428--G14 and NGC~7212. The detailed
relation between line brightness and excitation will be influenced by
factors additional to density, inlcuding the size of the cloud, the
local intensity of the ionizing radiation, abundances, and whether the
clouds are matter- or ionization-bounded. In addition, the inclusion
of some of the [\ion{N}{2}]$\lambda\lambda$6548,6583 lines in our
H$\alpha$ images complicates the interpretation of these
diagrams. It is, thus, not surprising that simple correlations are not
seen in Fig.~\ref{f14}.

Finally, it is interesting to consider the jets and NLR of Seyfert
galaxies in the broader context of AGN physics. As pointed out before,
some of our galaxies (e.g. Mrk~34 and Mrk~573) have [\ion{O}{3}]
luminosities comparable to radio-quiet, low-redshift quasars. For such
objects, we are discussing a regime in which the quasar and Seyfert
classifications overlap. One important feature of quasars
is, however, that they come in two classes: radio-loud and
radio-quiet. Using the well known [\ion{O}{3}]-to-radio correlations
(e.g.~Miller, Rawlings, \& Saunders 1993; Falcke, Malkan, \&
Biermann~1995; Falcke 1996), it is clear that the Seyferts in our sample belong to
the radio-quiet class of AGN, yet they not only show jets, but these
jets are also kinematically important for the emission-line gas. The
question of whether radio-quiet quasars also have jets has been
debated for a while.  Kellerman et al.~(1994) found a number of
radio-quiet quasars which have a double radio structure and Falcke,
Patnaik, \& Sherwood (1996a) found evidence for relativistic jets in
radio-quiet quasars. The jets we find in Mrk~34 and Mrk~573 would in
fact resemble some of the double structures seen by Kellerman et
al.~(1994) if placed at a larger distance and observed with lower
sensitivity.

\section{Conclusions}
Our results have shown that observations of jets and NLRs in Seyfert
galaxies allow a detailed look into the physics of the circumnuclear
regions of AGN, with much better linear resolution than is possible
for most of the more luminous AGN. Further studies of the NLR of
Seyferts should take account of the fact that Seyfert jets leave their
trace on the ISM. High (spatial) resolution spectroscopy, as it is
possible now with STIS on-board the HST, should provide quantitative
results on the kinematics, excitations, and densities involved in
these jet-ISM interaction regions and, in this way, give us important
information not only on the kinematics of the NLR, but also on the
kinematics and dynamics of low-power radio jets. Those observations
may also settle the exact nature of the excitation of the NLR. Even
though the radio jets substantially influence the NLR, it is still
unclear whether the dominant source of ionizing photons is the
compact, hidden Seyfert 1 nucleus or hot gas created in the jet-driven
shocks.

\acknowledgements We thank the STScI staff -- J. Biretta, M. McMaster,
and K. Rudloff -- for their support in the LRF calibrations. We are
are very grateful to R.W.~Argyle (Royal Greenwich Observatory) for
providing us with astrometric positions for the field stars near
NGC~4388 and NGC~7212 prior to publication and to R.R.J. Antonucci for
informations on his VLA observations of Mrk~1210 together with
R. Barvainis. We also thank R. Pogge for refereeing the paper and the
staff at NRAO in Socorro---especially J.S.~Ulvestad---for their
support. This research was supported by the Space Telescope Science
Institute under grant GO5411, by NASA under grants NAGW-3268,
NAGW4700, NAG8-1027, and in part by the DFG (Fa 358/1-1), and has made
use of the NASA Extragalactic Database (NED) which is operated by
Caltech/JPL. The National Radio Astronomy Observatory is a facility of
the National Science Foundation, operated under a cooperative
agreement by Associated Universities, Inc. Part of this work was
performed by the Jet Propulsion Laboratory, California Institute of
Technology under a contract with NASA.

\onecolumn

\begin{figure*}
\centerline{\fbox{See f4.gif}}
\caption[]{\label{f4}As Fig.~\ref{f3} for NGC~4388.
}\end{figure*}

\begin{figure*}
\centerline{\fbox{See f5.gif}}
\caption[]{\label{f5}As Fig.~\ref{f1} for NGC~7212.
}\end{figure*}

\begin{figure*}
\centerline{\fbox{See f13.gif}}
\caption[]{\label{f14}
The four panels show the logarithm of the emission-line brightness
versus the [\ion{O}{3}]-to-H$\alpha$+[\ion{N}{2}] ratio of each pixel
in our images of Mrk~573 (upper left), ESO~428$-$G14 (upper right),
NGC~4388 (lower left), and NGC~7212 (lower right). The fluxes per
pixel for the ordinate were taken from the [\ion{O}{3}] image for
ESO~428$-$G14 (because H$\alpha$+[\ion{N}{2}] was taken on the PC) and
from the H$\alpha$+[\ion{N}{2}] image for the rest.
}\end{figure*}

\end{document}